\documentclass[aps,prb,twocolumn,10pt]{revtex4-1}

\setcitestyle{numbers,square}

\def\arxiv{1}

\ifx\arxiv\undefined
    \newcommand{\secao}{\textit} 
    \newcommand{\arxivSM}[2]{#1} 
\else
    \newcommand{\secao}{\section} 
    \newcommand{\arxivSM}[2]{#2} 
\fi

\usepackage{xr}

\usepackage{physics}
\usepackage[english]{babel}
\usepackage[utf8]{inputenc}
\usepackage{amssymb}
\usepackage{amsmath}
\usepackage{multirow}

\usepackage[pdftex]{graphicx}
\usepackage{xspace}
\usepackage{dsfont}
\usepackage{bm}
\usepackage{braket}
\usepackage[normalem]{ulem}
\usepackage{hyperref}
\hypersetup{
    colorlinks,%
    citecolor=blue,%
    filecolor=blue,%
    linkcolor=blue,%
    urlcolor=blue
}

\newcommand{\bi}{Bi$_2$Se$_3$\xspace}
\newcommand{\mbi}{Bi$_2$Se$_3$/Ti\xspace}
\providecommand{\BiSe}{B\lowercase{i}$_2$S\lowercase{e}$_3$\xspace}

\providecommand{\ket}[1]{\vert #1\rangle} 

\providecommand{\M}{\mathcal{M}}

\providecommand{\Pc}{\mathcal{P}}
\providecommand{\Pt}{\mathcal{P}'}

\providecommand{\Q}{\mathcal{Q}}

\usepackage[usenames, dvipsnames]{color}

\begin{document}

\title{Confinement and fermion doubling problem in Dirac-like Hamiltonians}
\author{B. Messias de Resende}
\author{F. Crasto de Lima} 
\author{R. H. Miwa}
\author{E. Vernek}
\author{G. J. Ferreira}
\affiliation {Instituto de F\'{i}sica, Universidade Federal de Uberl\^{a}ndia, Uberl\^{a}ndia, MG 38400-902, Brazil}
\date{\today}

\begin{abstract}
We investigate the interplay between confinement and the fermion doubling problem in Dirac-like Hamiltonians. Individually, both features are well known. First, simple electrostatic gates do not confine electrons due to the Klein tunneling. Second, a typical lattice discretization of the first-order derivative $k \rightarrow -i\partial_x$ skips the central point and allow spurious low-energy, highly oscillating solutions known as fermion doublers. While a no-go theorem states that the doublers cannot be eliminated without artificially breaking a symmetry, here we show that the symmetry broken by the Wilson's mass approach is equivalent to the enforcement of hard-wall boundary conditions, thus making the no-go theorem irrelevant when confinement is foreseen. We illustrate our arguments by calculating the following: (i) the band structure and transport properties across thin films of the topological insulator \BiSe, for which we use \textit{ab-initio} density functional theory calculations to justify the model; and (ii) the band structure of zigzag graphene nanoribbons.
\end{abstract}
\maketitle

\arxivSM{}{\secao{Introduction}}

Topological insulators (TIs) constitute a class of materials that exhibit the ubiquitous property of being an insulator in their bulk, while presenting metallic states on their edges or surfaces \cite{PhysRevLett.96.106802, Konig766, RevModPhys.82.3045, RevModPhys.83.1057}. The key ingredient for the underlying physics of the TIs is a strong spin-orbit interaction, which generically leads to a Dirac-like spectrum. At low energy, the effective Hamiltonians for the edge/surface states are linear in the momentum, yielding a massless Dirac spectrum. The resulting helical band structure is topologically protected against backscattering, thus providing perfect conducting channels that are potentially useful for future electronic devices \cite{PhysRevApplied.2.054010,Thomas2016}, quantum computation \cite{PhysRevB.88.085316}, and optical applications \cite{Yeatse1500640}.

\begin{figure}[ht!]
  \centering
  \includegraphics[width=\columnwidth,keepaspectratio]{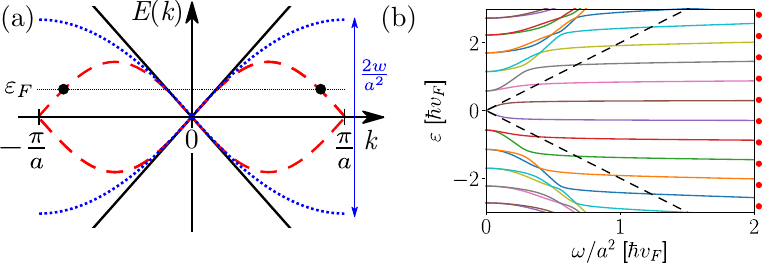}
  \caption{(a) In a discrete lattice with spacing $a$, the linear dispersion $\varepsilon = \pm \hbar v_F k$ of the continuous model (solid black line) is replaced by $\varepsilon = \pm \hbar v_F (2a)^{-1}\sin(ka)$ (dashed red line), yielding the ``doublers'' at the Fermi energy $\varepsilon_F$ (black dots). A finite Wilson's mass term $\propto w k^2$ eliminates the doublers by opening a gap at $k = \pm \pi/a$ (dotted blue line). (b) Quantized energies of the linear spectrum due to hard-wall confinement as a function of the Wilson mass $w$. The numerical solutions (solid lines) approach the exact solutions (red dots) in the range $|\varepsilon| < 2w/a^2$ (black dashed lines). For $w \rightarrow 0$ the numerical solutions merge to form the doublers.}
  \label{Fig1}
\end{figure}

The numerical approach to investigate the properties (e.g., transport, dynamics, confinement) \cite{PhysRevB.84.073109, 0953-8984-9-2-014, hao2011topological} of these systems often requires a lattice discretization of the Hamiltonian. Unfortunately, standard finite difference descriptions of the first-order derivatives of linear in momentum $\hbar k$ Hamiltonians are infected by the fermion doubling problem (FDP). This yields spurious low-energy states, as exemplified in Fig.~\ref{Fig1}(a). Even though the energy dispersion is well described by the discrete Hamiltonian at small $k$, the doublers appearing for large $k$ will affect the transport and dynamics of the system. There are many ways to eliminate the doublers \cite{DeGrand2007, chandrasekharan2004introduction}, e.g., staggered fermions \cite{Susskind1977Staggered, stacey82, finitedifference2008, hernandez2012finite}, Wilson's mass \cite{wilson1974confinement, Kogut1975, wilsonfermionsoptical, zhou2016LatticeModel}, non-local discretizations \cite{Svetitsky1980SLAC, Quinn1986SLAC, SLAC12}, and extra artificial dimensions \cite{kaplan1992method, Kaplan2012Spacetime, creutz1994Talk, creutz1994surface}. Each of them presents its own advantages and disadvantages. There is, however, a common and seemingly unsolvable  problem: As required by the Nielsen-Ninomiya theorem (NNT) \cite{NIELSEN1981219, *nielsen1981absence, Karsten1981}, all these approaches introduce a symmetry breaking or nonlocality.

The $k$-linear models also display the Klein tunneling ``paradox'' \cite{Klein1929, katsnelson2006chiral}, which states that simple electrostatic barriers are transparent and cannot confine massless electrons. This is a consequence of the constant Fermi velocity of the linear dispersion, which allows perfect matching of the injected and transmitted waves. Consequently, to attain confinement, one needs to either open a gap by breaking a symmetry in the outer region \cite{berry1987neutrino, Alonso1997Boundary, mccann2004symmetry, Fabian2014TIDots, ferreira2013magnetically, yeats2017local, zhou2016LatticeModel}, or invoke finite-size effects \cite{Ferreira2011LowBiasNDR}.

In this Rapid Communication we ask whether it is possible to eliminate the FDP in a finite system by breaking the same symmetry that provides the confinement. The answer is yes. The short argument is as follows: Since the symmetry is already broken by the confinement, there is no harm in introducing a Wilson's mass that breaks the same symmetry. More interestingly, here we show that the Wilson's mass not only eliminates the doublers, but also defines the type of hard-wall confinement that is imposed by vanishing boundary conditions. To present this argument, we start with a simple unidimensional model that captures its essence. Here, we solve the linear Hamiltonian with vanishing flux hard-wall boundaries \cite{mccann2004symmetry, ferreira2013magnetically}, and compare it with the solutions obtained by introducing a parabolic Wilson's mass term $\propto wk^2$ and vanishing wave-function hard-wall boundaries. We find that the solutions match for a small, but finite, Wilson's mass $w$, while for $w \rightarrow 0$, one recovers the spurious doublers.

Next, we discuss the surface states of the three-dimensional (3D) topological insulator \BiSe as a prototype model to illustrate our findings. Here, we consider two different forms of the Wilson's mass term to show that it can either break time-reversal symmetry (TRS), as in Refs.~\cite{ferreira2013magnetically, zhou2016LatticeModel, yeats2017local}, or a sublattice chiral symmetry. Its consequences for the energy levels and degeneracies of a quantum dot, and the transport properties across a ribbon are discussed. We use a modified version of the effective Hamiltonian for the \BiSe from Refs.~\onlinecite{zhang2009topological, shan2010effective} fitted to first-principles calculations from the VASP code 
\arxivSM{
    \cite{vasp1, SM}.
}{
    \cite{vasp1} (details in Appendix \ref{ap:dft}).
}
The effective model is then implemented numerically using the KWANT code \cite{kwant}. Additionally, we briefly present the case of zigzag graphene nanoribbons, which is a challenging case for effective models \cite{hernandez2012finite}.

\secao{Fermion doubling.}
To establish our arguments, let us first consider a simple one-dimensional Dirac-like model given by the Hamiltonian $H_\xi = \hbar v_F \M_\xi k$, where $v_F$ is the Fermi velocity, $k$ is the momentum along a generic coordinate $\xi$, and $\M_\xi$ is a unitary Hermitian matrix. The exact energy spectrum of $H_\xi$ is $\varepsilon = \pm \hbar v_F k$, which is the Dirac cone illustrated in Fig.~\ref{Fig1}(a). However, if one desires to find the spectrum numerically via finite differences, the momentum $\hbar k = -i\hbar\partial_{\xi}$ takes a discrete form. To keep $H$ Hermitian, one typically chooses the symmetric finite difference approach, leading to an expression that skips the central point, i.e., $\partial_\xi\psi(\xi_j) \approx [\psi(\xi_{j+1}) - \psi(\xi_{j-1})]/2a$, where the integer $j$ labels the points in the discrete lattice of spacing $a$. Consequently, it allows for low-energy, highly oscillating states of a topological origin \cite{Karsten1981, NIELSEN1981219, nielsen1981absence}, thus yielding the doublers shown in Fig.~\ref{Fig1}(a).

\secao{Wilson's mass.}
Here, we choose the Wilson's mass approach \cite{wilson1974confinement, Kogut1975, wilsonfermionsoptical, zhou2016LatticeModel} to eliminate the doublers. The idea is to introduce a parabolic correction $H_{\rm W} = w\M_ck^2$ to $H_\xi \rightarrow H_\xi + H_{\rm W}$. For small $k$, the linear terms dominate and $H_{\rm W}$ does not significantly affect the band structure. However, this term eliminates the doublers as its discretization couples all three points $j$, and $j\pm 1$, thus opening a gap $2w/a^2$ at $k=\pm\pi/a$, as shown in Fig.~\ref{Fig1}(a). The penalty for using $H_{\rm W}$ is that it breaks a chiral symmetry of the linear $H_\xi$. Next, we argue that this penalty is irrelevant if one chooses $\M_c$ to be the same unitary matrix that the defines the hard-wall confinement.

\secao{Hard-wall boundary conditions.}
The hard-wall boundary condition for $H_\xi$ (without $H_{\rm W}$) is imposed by the limit $\alpha\rightarrow\infty$ of the confining potential $H_C = \alpha \M_c\Theta(|\xi|-\xi_0)$, where $\Theta(\xi)$ is the Heaviside step function defining the walls at $\xi=\pm\xi_0$. The unitary matrix $\M_c$ must break a symmetry of $H_\xi$ to open a gap $2\alpha$ in the outer region ($|\xi| > \xi_0$). At the interface, the spinor is discontinuous \cite{berry1987neutrino,Alonso1997Boundary}, and integrating $H_\xi\psi = \varepsilon\psi$ across the interface we obtain the boundary condition \cite{mccann2004symmetry,ferreira2013magnetically}
\begin{equation}
    \Big(\pm i \M_\xi + \M_c\Big)\psi(\xi_0) = 0,
    \label{eq:BC}
\end{equation}
where the matrices $(\pm i\M_\xi+\M_c)$ are singular, thus allowing nontrivial solutions. Notice that we have used the same matrix $\M_c$ to define here this boundary condition for linear $H_\xi$, and above to introduce the Wilson's mass parabolic term $H_{\rm W}$. This assures that both approaches will break the same symmetry of $H_\xi$.

In contrast to Eq.~\eqref{eq:BC}, the Wilson's mass model $H_\xi+H_{\rm W}$ allows trivial vanishing boundary conditions $\psi(\pm \xi_0) = 0$. Therefore, now we have two different approaches to apply a hard-wall confinement. Moreover, our choice of a simple $H_\xi$ allows for analytical solutions (up to a transcendental equation) for the boundary condition from Eq.~\eqref{eq:BC}, thus avoiding the discretization and the FDP all together. In Fig.~\ref{Fig1}(b) we compare these solutions with a numerical finite difference model for $H_\xi+H_{\rm W}$ with vanishing boundary conditions as a function of the Wilson's mass $w$. For $w \rightarrow 0$, pairs of eigenstates merge to form the degenerate doublers, but are split for finite $w$. An appropriate value for $w$ can be chosen such that the energy window of interest lies within the Wilson's mass gap $2w/a^2$ [see Fig.~\ref{Fig1}(a)], and preserves the dominance of the linear term over the parabolic correction, which yields $\frac{1}{2}a^2|\varepsilon| < w < (\hbar v_F)^2/|\varepsilon|$ 
\arxivSM{\cite{SM}}{(see Appendix \ref{ap:massrange})}.

\secao{\BiSe thin films.}
Next, we apply our approach to model thin films of the topological insulator \BiSe.
Bulk \BiSe is composed of van der Waals interacting quintuplelayers (QLs). Each QL is formed by an alternation of covalent bonded hexagonal monolayers of Se-Bi-Se-Bi-Se. The effective Hamiltonian for both the bulk and its surface states are well known \cite{zhang2009topological, shan2010effective, ferreira2013magnetically}. Here, we choose to write it on the basis of surface states of semi-infinite solutions from the top (T) and bottom (B) surfaces, i.e., $\{\varphi_{T\uparrow}(\bm{r}), \varphi_{T\downarrow}(\bm{r}), \varphi_{B\uparrow}(\bm{r}), \varphi_{B\downarrow}(\bm{r})\}$, where $\{\uparrow, \downarrow\}$ refers to the spin along $z$. Up to linear order in $\bm{k} = (k_x, k_y)$ the Hamiltonian reads 
\arxivSM{\cite{SM}}{(see Appendix \ref{ap:model})}
\begin{equation}
    H = \varepsilon_0 + \hbar v_F(k_x\gamma_{3y}-k_y\gamma_{3x}) + F\gamma_{30} + \Delta\gamma_{10} + B\gamma_{0z},
    \label{eq:H2D}
\end{equation}
where $\gamma_{ij} = \tau_i\otimes\sigma_j$, $\bm{\sigma}$ and $\bm{\tau}$ are $\mathfrak{su}(2)$ operators acting on spin and surface subspaces, $\varepsilon_0$ is the energy reference, $F$ represents the intensity of a structural inversion asymmetry (SIA) field, $\Delta$ is the hybridization coupling between the surfaces, and $B$ is a generic Zeeman field. 
We extract these parameters from DFT simulations \arxivSM{\cite{Forster2015GW, SM}}{\cite{Forster2015GW} (see Appendix \ref{ap:dft})}. For a pristine \BiSe stacking of seven QLs, the Dirac bands are well defined and we find $v_F = 479$~nm/ps \cite{natureZhang2010}, $\varepsilon_0 = -12$~meV, $F=\Delta\approx 0$. Additionally, in the \arxivSM{Supplemental Material}{Appendix \ref{ap:dft}} we analyze the band structure of \BiSe contacted by a Ti metallic lead. At this interface, a charge transfer yields a bias field $F \approx 95$~meV and a shift of the Dirac cones $\varepsilon_0 \approx -150$~meV. Moreover, these are coupled to metallic bands near the Fermi level, which will allow us to use the wide-band approximation later on.


A finite $F$ splits the Dirac cones from the top and bottom surfaces without opening a gap, while $\Delta$ opens a gap hybridizing the surfaces, and $B$ opens a gap by breaking TRS. Therefore, there are two possible Wilson mass terms that can be added to $H$ to eliminate the doublers and define the types of hard-wall boundary conditions. These are
\begin{equation}
    H_B = m_B \dfrac{a^2}{4} k^2 \gamma_{0z}, \;\; \text{ and } \;\; H_\Delta = m_\Delta \dfrac{a^2}{4} k^2 \gamma_{10}.
    \label{eq:mass}
\end{equation}
Hereafter we will refer to $m_B$ and $m_\Delta$ as the Wilson masses for a $B$-type and $\Delta$-type hard-wall confinements. These break the same symmetries as $B$ and $\Delta$. Similarly to the range of $w$ above, the appropriate range for $m_{B/\Delta}$ is $|\varepsilon|/2 < |m_{B/\Delta}| < \left(2\hbar v_F/a\right)^2/|\varepsilon|$ 
\arxivSM{
    \cite{SM}.
}{
    (see Appendix \ref{ap:massrange}).
}

\secao{Chiral symmetries.}
A chiral symmetry \cite{PhysRevB.55.1142, bernard2002classification, Schnyder2008PRBClassification} is defined by an operator $\Pc$ that anticommutes with $H$. Consequently, it assures that for every eigenstate of $H$ with energy $\varepsilon$, there is a chiral partner with energy $-\varepsilon$. For the \BiSe $H$ above, we find four candidate operators for chiral symmetries that obey 
\begin{align}
 \{\Pc_{0z}, H-\varepsilon_0\} &= 2B + 2\Delta\gamma_{1z} + 2F\gamma_{3z},\\
 \{\Pc_{10}, H-\varepsilon_0\} &= 2B\gamma_{1z} + 2\Delta,\\
 \{\Pc_{20}, H-\varepsilon_0\} &= 2B\gamma_{2z},\\
 \{\Pc_{3z}, H-\varepsilon_0\} &= 2B\gamma_{30} + 2F\gamma_{0z},
\end{align}
where $\Pc_{ij}\equiv\gamma_{ij}$. For simplicity, we omit the Wilson masses $m_B$ and $m_\Delta$, but their contributions follow the $B$ and $\Delta$ terms above. In accordance with the NNT \cite{NIELSEN1981219, nielsen1981absence, Karsten1981}, a finite $m_B$ or $m_\Delta$ breaks some chiral symmetries. Particularly, $m_B\neq 0$ breaks them all. However, we find that combining the $\Pc$ operators above with the TRS operator $\mathcal{T} = -i\gamma_{0y}\mathcal{K}$ ($\mathcal{K}$ is complex conjugation) as $\Pt_j = \Pc_j\mathcal{T}$, one obtains similar anticommutation relations independent of $B$ 
\arxivSM{
    \cite{SM}.
}{
    (see Appendix \ref{ap:chiral}).
}

\begin{figure}[ht!]
  \centering
  \includegraphics[width=\columnwidth,keepaspectratio]{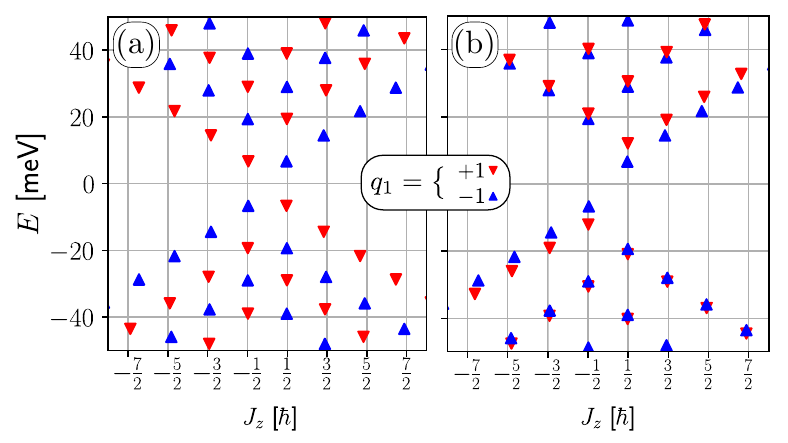}
  \caption{Spectrum of a circular \BiSe thin-film quantum dot as a function of the total angular momentum $j_z$. The up- (blue) and down-pointing (red) triangles refer to the chiral charge $q_{1z}=\pm 1$. (a) For the $\Delta$-type confinement both the chirality ($\varepsilon_{j_z, -q_{1z}} = -\varepsilon_{j_z, q_{1z}}$) and TRS ($\varepsilon_{-j_z, -q_{1z}} = \varepsilon_{j_z, q_{1z}}$) are preserved. (b) The $B$-type confinement breaks both chiral and TRS, but preserves their product ($\varepsilon_{-j_z, q_{1z}} = -\varepsilon_{j_z, q_{1z}}$).}
  \label{Fig2}
\end{figure}

\secao{Circular dot.}
To illustrate the $\Delta$- and $B$-type confining potentials and the chiral symmetries, we consider a circular quantum dot of radius $R$ on a \BiSe thin film modeled by $H$ in Eq.~\eqref{eq:H2D} and the Wilson masses in Eq.~\eqref{eq:mass}. For this geometry, the $z$-component $J_z = L_z + S_z$ of the total angular momentum is conserved \cite{ferreira2013magnetically}, which allows us to label the states by its eigenvalues $j_z = (m+\frac{1}{2})\hbar$, where $m$ is an integer. The discrete spectrum of this quantum dot for $\Delta$-type ($m_\Delta = -500$~meV) and $B$-type ($m_B = -500$~meV) confining potentials are shown in Fig.~\ref{Fig2} as a function of $j_z$. For simplicity, in both cases, $\varepsilon_0 = F = B = 0$, $\Delta = -5$~meV, and $v_F = 479$~nm/ps. The eigenvalues are obtained using a square lattice in which the sites are connected only for $r \leq R = 50$~nm, and $N=100$ sites along the diagonal.

For the $\Delta$-type confinement shown in Fig. \ref{Fig2}(a), $\Pc_{20}$ and $\Pc_{3z}$ are chiral symmetries. These combine to define the chiral charge $\Q_{1z} = -i\Pc_{20}\Pc_{3z} = \gamma_{1z}$, such that $[\Q_{1z}, H] = 0$. Together with $j_z$, the eigenvalues $q_{1z} = \pm 1$ of $\Q_{1z}$ are used to label the eigenenergies as $\varepsilon_{n,j_z,q_{1z}}$, where $n$ is an extra index that labels the different solutions with the same $j_z$ and $q_{1z}$. Since $\{\Pc, H\} = 0$, $[\Pc, J_z] = 0$ and $\{\Pc_j, \Q_{1z}\} = 0$, every state with energy $\varepsilon_{n,j_z,q_{1z}}$ has a chiral partner with energy $\varepsilon_{n,j_z,-q_{1z}} = -\varepsilon_{n,j_z,q_{1z}}$. Similarly, the TRS produces the Kramer partners with energies $\varepsilon_{n,-j_z,-q_{1z}} = \varepsilon_{n,j_z,q_{1z}}$. Combined, these two symmetries produce the X-shaped spectrum of Fig.~\ref{Fig2}(a). In contrast, Fig.~\ref{Fig2}(b) shows the spectrum for the $B$-type confinement, for which the chiral symmetries $\Pc$ and TRS are broken. Here, the time-reversal chiralities $\Pt_{20}$ and $\Pt_{3z}$ are preserved. These combine to give the same chiral charge $\Q_{1z}$. However, now $\{\Pt, J_z\} = 0$ and $[\Pt, \Q_{1z}] = 0$. Consequently, a state with energy $\varepsilon_{n,j_z,q_{1z}}$ has a time-reversal chiral partner with energy $\varepsilon_{n,-j_z,q_{1z}} = -\varepsilon_{n,j_z,q_{1z}}$, yielding the single linear branch and the shifted bands in Fig.~\ref{Fig2}(b).
The agreement between these exact relations and the numerical results in Fig.~\ref{Fig2} show that our approach does eliminate the doublers without any harm to the chiralities that remain in the presence of confinement.

\begin{figure}[ht!]
  \centering
  \includegraphics[width=\columnwidth,keepaspectratio]{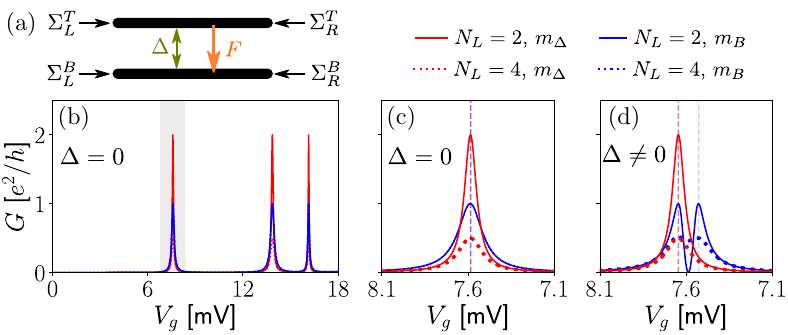}
  \caption{(a) Illustration of the top and bottom \BiSe surfaces coupled by $\Delta$, biased by $F$, and contacted by effective leads introduced by the self-energies $\Sigma_\ell^j$.
  (b) Conductance vs $V_g$ for $\Delta = 0$. The $V_g \sim 7.6$~mV peak is shown in detail in (c) for $\Delta=0$, and (d) for $\Delta = 0.1$~meV. Red (blue) lines correspond to $\Delta$-type ($B$-type) confinement, and solid (dashed) lines refer to the $N_L = 2$ $(4)$ terminal model.}
  \label{Fig4}
\end{figure}

\secao{Conductance across a ribbon device.}
As another example of our main result, let us now calculate the conductance across the \BiSe surface. We consider a geometry that was recently realized experimentally \cite{SRSungjae2016}, where the leads are contacted with metal electrodes, while the scattering region is pristine \BiSe. The conductance peaks reflect the degeneracy of the states, which are directly affected by the symmetry breaking discussed previously. In the leads, the hybridization between the topmost QLs \BiSe and the metal \cite{JPCMMiwa2017} puts the chemical potential within an energy window composed of \BiSe surface and Ti states \arxivSM{\cite{SM}}{(see Appendix \ref{ap:dft})}.
Therefore, we can judiciously assume that the effect of the leads is essentially to broaden the discrete Fabry-Perot resonances in the confined central region. Within this simplified description, we introduce the self-energies $\Sigma^{j}_{\ell}(E)$  with $\ell = L,R$ (for left and right) and $j=T,B$ (for top and bottom), which in the wide-band limit are $\Sigma^{j}_{\ell}(E) =-i\bar\Sigma^j_{\ell}(E_F)\Theta(D-|E-E_F|)$; see Fig.~\ref{Fig4}(a). Here, $\bar \Sigma^j_{\ell}(E_F)$ is a real quantity giving the broadening of the sites interfacing the $\ell$th TI lead, and $D$ is some suitable cutoff energy. This rather crude simplification is very suitable for numerical simulations of realistically sizable systems. Nonetheless, it gives  qualitatively plausible results for the conductance as compared to those obtained with a complete model. 

For the scattering region we consider square surfaces of pristine \BiSe of side $W = 100$~nm, discretized into a $20\times 20$ site grid. A two-terminal case ($N_L=2$) is built with the top left and right ($\ell = L$ and $R$) leads with symmetric broadening $\bar\Sigma^T_\ell=1$~meV. Moreover, despite the reduced coupling to the metal contact, a four-terminal ($N_L=4$) case is also considered with $\bar\Sigma^B_\ell=1$~meV. Figure~\ref{Fig4}(b) shows the conductance $G$ versus gate voltage $V_g$ for $\Delta=0$, $N_L = 2(4)$ as solid (dashed) lines, and both $\Delta$-type (blue lines, $m_\Delta=-100$~meV) and $B$-type (red lines, $m_B=-100$~meV) confinements. Figures \ref{Fig4}(c) and \ref{Fig4}(d) zoom to show details of the $V_g \sim 7.6$~mV peaks. For the $N_L=2$ case with the $\Delta$-type confinement, we observe that the peaks reach $2G_0$, where the factor 2 results from the TR pair of degenerate states (symmetric and antisymmetric combinations of the top and bottom surface states) that contribute as independent conducting channels. These peaks are not substantially affected by the presence of small $\Delta\neq 0$, as we see in Fig.~\ref{Fig4}(d). In contrast, for the $B$-type confinement the conductance peaks reach only $G_0$ in Fig.~\ref{Fig4}, for $\Delta=0$. Indeed, here TRS is broken and one would already expect a single conducting channel. More interestingly, a finite $\Delta=0.1$~meV splits this peak, showing $G=0$ in the middle. This can be understood in terms of the $\Pt$ chiral symmetries and the conserved chiral charge $\Q_1$, which assures that for $\Delta=0$, every state located in one surface has a degenerate partner in the other surface 
\arxivSM{\cite{SM}}{(see Appendix \ref{S:localization})}.
A finite $\Delta$ couples these partners, producing two coherent channels that interfere destructively ($G=0$) for some particular value of $V_g$. For $N_L=4$, the conductance is still calculated between the top terminals (dashed lines in Fig.~\ref{Fig4}). Overall, this yields a decrease of $G$ whenever the top and bottom surfaces are coupled ($\Delta$-type confinement or $\Delta\neq 0$). In this situation, the channels involving the bottom surface states are broadened by the bottom contacts, therefore they act as incoherent channels, destroying the perfect inference. 

\begin{figure}[ht!]
  \centering
  \includegraphics[width=\columnwidth,keepaspectratio]{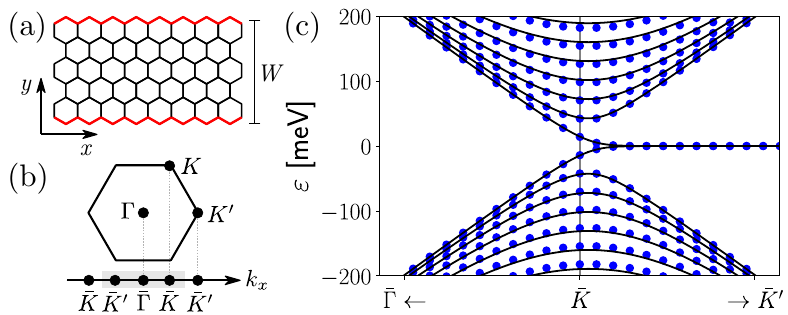}
  \caption{(a) A graphene nanoribbon with zigzag terminations and width $W$. (b) The first Brillouin zone of bulk graphene and its projection (shaded area) along the nanoribbon's $k_x$. (c) Band structure around $\bar{K}$ for a $W \approx 71$~nm ribbon comparing the analytical solution (black solid lines) and our numerical approach with a Wilson mass (blue dots).}
  \label{graphene}
\end{figure}

\secao{Zigzag graphene nanoribbon.} 
As a final application of our proposal, we present the band structure of a zigzag graphene nanoribbon around its $K$ point in Fig.~\ref{graphene}. This is a particularly interesting case as it allows us to compare the numerical results directly with well known analytical solutions \cite{Nakada1996GrapheneRibbons,BreyFertig2006Graphene, BreyFertig2006Dirac, CastroNeto2009Review}, which are shown as black solid lines in Fig.~\ref{graphene}(c). For the numerical approach we start with graphene's effective model around $K$, $H_K = \hbar v_F \bm{\sigma}\cdot(\bm{k}-\bm{K})$, where $\bm{k}$ is measured from the origin at $\bar{\Gamma}$ in Fig.~\ref{graphene}(b), and $k_y \rightarrow -i\partial_y$ is discretized into $N=100$ sites. Around $K'$ one obtains $H_{K'}$ replacing $\sigma_y \rightarrow -\sigma_y$ and $\bm{K} \rightarrow \bm{K}'$, which compose our block-diagonal $H_0 = H_K \oplus H_{K'}$. To regularize the boundary conditions for the zigzag nanoribbon we consider a Wilson mass term $H_Z = m_z\frac{a^2}{4}k_y^2 (\tau_x\otimes\sigma_y)$, where $\tau_x$ couples the $K$ and $K'$ subspaces, and $m_z$ is chosen within the range set by the inequalities discussed previously. The agreement between the numerical band structure and the exact solution shown in Fig.~\ref{graphene}(c) is patent, which illustrates the effectiveness of our approach.

\secao{Conclusions.}
We have shown that the Wilson mass not only eliminates the doublers in Dirac-like Hamiltonians, but also allows us to control the hard-wall boundary conditions. 
This contrasts with the high-energy physics, where neither the broken symmetry nor confinement are desirable. Therefore, for confined solid state systems, the NNT is easily bypassed. Interestingly, these effects were overlooked in models that already include the parabolic terms \cite{bernevig2006quantum, hao2011topological, michetti2012helical, zhou2016LatticeModel}. Indeed, in the Bernevig-Hughes-Zhang (BHZ) model \cite{bernevig2006quantum}, for instance, the term $-B k^2 \sigma_z$ plays the role of the Wilson mass, with the Pauli matrix $\sigma_z$ acting on the $E_1/H_1$ subspace, yielding a Dirac mass-type hard wall~\cite{michetti2012helical}.
In contrast, graphene models are usually restricted to the linear terms, which limits its use. Here, we have seen that a zigzag termination can be well modeled by incorporating an appropriate Wilson's mass. For the armchair case, one can directly combine Ref.~\cite{mccann2004symmetry} with our approach.

Applying our model to model \BiSe quantum dots, we have shown that numerical results satisfy all symmetry constraints that are compatible with the chosen type of confinement. Particularly, the $\Delta$-type confinement is compatible with thin films \cite{hao2011topological}, yielding noninteracting conductance peaks $G = 2\;e^2/h$, which is a necessary ingredient for the Kondo regime suggested in Ref.~\onlinecite{SRSungjae2016}. As a final remark, notice that Ref.~\onlinecite{zhou2016LatticeModel} considers only a $B$-type mass, which breaks TRS, and the confinement properties are not discussed. Therefore, our model generalizes and improves their results.

The authors acknowledge the financial support from the Brazilian Agencies CNPq, CAPES, and FAPEMIG.

\arxivSM{}{\appendix 

\section{DFT Model and Results}
\label{ap:dft}


The calculations were performed based on the DFT approach, as implemented 
in the VASP code \cite{vasp1}. The exchange correlation term was described using 
the GGA approach in the form proposed by Perdew, Burke and Ernzerhof 
(PBE) \cite{PBE}. The Kohn-Sham orbitals are expanded in a plane wave basis set 
with an energy cutoff of 400 eV. The 2D Brillouin Zone (BZ) is sampled according 
to the Monkhorst-Pack method \cite{Monkhorst1976BZpoints}, using a 8$\times$8$\times$1 
mesh. The electron-ion interactions are taken into account using the Projector 
Augmented Wave (PAW) method \cite{Blochl1994PAW}. All geometries have been 
relaxed until atomic forces were lower than $0.025$\,eV/{\AA}. The van der Waals 
interactions (vdW-DF2 \cite{VDW-DF2}) were included  to correctly describe the 
system. In all cases, we have considered a vacuum region of  at least 
$24$\,{\AA} to avoid periodic-image interactions.


The metal/topological-insulator interface, {\mbi} \cite{SRSungjae2016}, was modeled by considering a Ti-{$\omega$} hexagonal slab of 14 atomic layers stacked over the hexagonal {\bi} (001) surface, which in turn is described by a slab composed by 7 quintuple layers (QLs), Fig.\,\ref{Fig3}(a). At low temperatures ($\sim 16$~mK) the $\omega$ phase of Ti is the most stable \cite{PRB80Mei}, with a lattice parameter $a = 4.57$~\AA~ \cite{PRB80Mei}. This remains true despite the $\sim 8\%$ compression needed to accommodate the lattice parameter $a = 4.21$~\AA~ of {\bi}.

\begin{figure}[ht]
  \centering
  \includegraphics[width=\columnwidth,keepaspectratio]{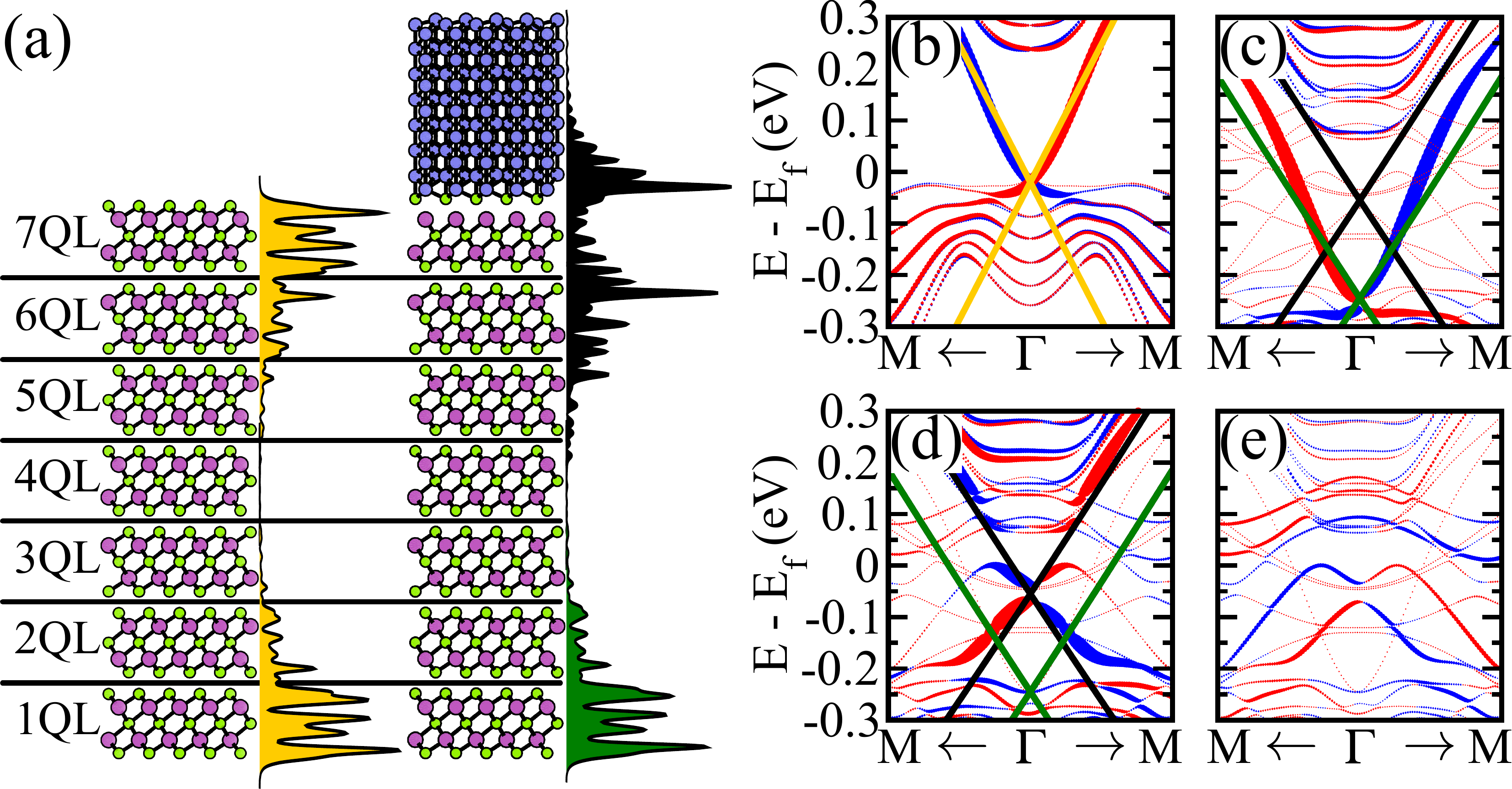}
  \caption{(a) Structural geometry and local DOS of {\bi} ({\mbi}) in the left (right) for $E$ near the Dirac points. Purple, green and blue circles are for Bi, Se and Ti atoms. Spin projected band structure ($|\langle \# QL|n,{\bf k} \rangle|^2$) with red (blue) dots labeling $\text{sign}({\langle S_y \rangle}) = +1$ ($-1$) for (b) {\bi} 7QL, (c) {\mbi} 1QL, (d) {\mbi} 6QL, (e) {\mbi} 7QL. The lines in (b)-(d) are fits to the surface states model, Eq.~\eqref{eq:H2D}, with the color code matching the DOS in panel (a).}
  \label{Fig3}
\end{figure}

The band structure of pristine \bi near $\Gamma$ shows the Dirac-like energy dispersion for states localized into the topmost QL (7QL), as seen in Fig.~\ref{Fig3}(b), for which we find $v_F = 479$~nm/ps, $\varepsilon_0 = -12$~meV, $F = \Delta \approx 0$. There, the dot size represents its localization into the surface, while the color refers to the spin up (blue) and down (red) projections, which evidences its helical nature. The bottommost QL is degenerate with this one, but with opposite spin projections. It is known that a better description of the surface states below the Fermi level for \BiSe requires a GW calculation \cite{Forster2015GW}, which is computationally expensive. However, despite \BiSe surface states below the Fermi energy been imperfectly described by DFT, the states above the Fermi energy are correctly described \cite{Forster2015GW}. In this way by fitting the {\it ab initio} band structure with the model Eq.\,\eqref{eq:H2D} for the states within $E>E_f$ we obtain a Fermi velocity of 479 nm/ps, which is only $\sim 6\%$ greater than the experimental observation \cite{natureZhang2010}. 

We find quite a different picture upon the formation of the \mbi interface.
At the equilibrium geometry, the topmost Se atoms of {\mbi} break the bond with Bi to attach covalently to the metal surface, Fig.\,\ref{Fig3}(a). We can see this scenario by the increase of the Bi-Se bond length $d_{\rm BiSe} = 2.86 \rightarrow 3.40$\,{\AA} at the interface region, followed by the formation Se-Ti chemical bonds, with $d_{\rm SeTi} = 2.63$\,{\AA} close to the sum of their covalent radius ($2.52$\,{\AA}). Such change in the {\bi} surface impacts the electronic structure of the topological states. 
Indeed,  charge transfers at the {\mbi} interface create a SIA field ($F=95$~meV, and $\varepsilon_0 = -150$~meV) that splits the degenerate Dirac bands. The states of the 1QL are mostly unaffected ($\Delta \approx 0$), Fig.~\ref{Fig3}(c), since their coupling to the metal is negligible. On the other hand, the states of the topmost QLs hybridize with the metal, Fig.~\ref{Fig3}(d), spreading-out the density of states (DOS) peak of the pristine case from the 7QL into the metal and the 6QL, Fig.~\ref{Fig3}(a). The spin-polarized Dirac dispersion is now located in 6QL, Fig.~\ref{Fig3}(d)-(e), but shows small gaps due to the hybridization with the metal bands. Nonetheless, we can still define a Dirac-like dispersion near the Fermi level, where the coupling to the metal is weak \cite{PRB82Zhao}. In this case we find $v_F = 338$~nm/ps.



In Fig.\,\ref{pdos1} we present the Projected Density of States (PDOS) for the leads \mbi. We can see that the 1QL orbitals (black line) maintain a quasi constant Density of States (DOS) in the range $-0.2 < E < 0.05$~eV, which is consistent with the preservation of the linearly dispersive band. Figure \ref{pdos1}(b) shows the PDOS close to the Fermi energy, where the PDOS from the 6QL and 7QL display smooth fluctuations. Such a picture of PDOS near the Fermi energy are in accordance with the wide band limit approximation used in the section {\it Conductance across a ribbon device} of the main text.

\begin{figure}[ht]
  \centering
  \includegraphics[width=0.9\columnwidth,keepaspectratio]{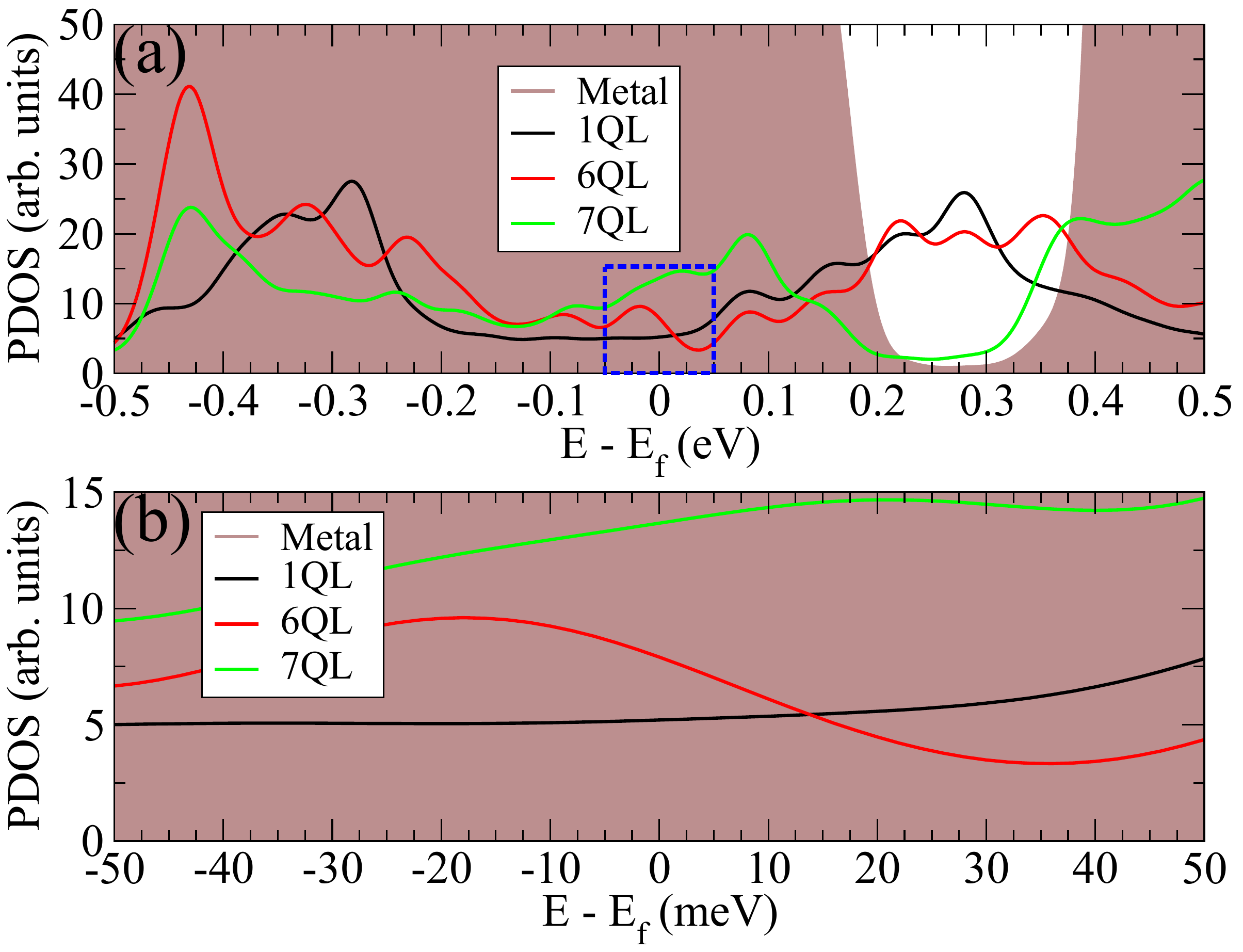}
  \caption{(a) Projected density of states of \mbi into the Metal (brown), 1QL (black), 6QL (red) and 7QL (green) orbitals. (b) Dashed blue rectangle of (a).}
  \label{pdos1}
\end{figure}


\section{Effective model for surface states}
\label{ap:model}

Near the $\Gamma$ point, the states around the Fermi energy of \BiSe are 
described by two Bi and Se hybridized $p_z$ orbitals $\{ \ket{p_z^{1+};\uparrow}, 
\ket{p_z^{2-};\uparrow}, \ket{p_z^{1+};\downarrow}, 
\ket{p_z^{2-};\downarrow} \}$, where the sign $\pm$ defines the parity 
under inversion, $\{1, 2\}$ correspond to the hybridization of two Bi and Se $p_z$ orbitals in the unit cell, and $\nu=\{\uparrow, \downarrow\}$ is the $z$ component of the 
spin. Onto this basis, the bulk Hamiltonian acquires the form 

\begin{equation}
    H = 
    \begin{pmatrix}
        C+M & A_1 k_z & 0 & A_2 k_- \\
        A_1 k_z & C-M & A_2 k_- & 0 \\
        0 & A_2 k_+ & C+M & A_1 k_z \\
        A_2 k_+ & 0 & A_1 k_z & C-M
    \end{pmatrix},
    \label{eq:HGamma}
\end{equation}
where $C$, $M$, $A_1$, $A_2$  are the symmetry allowed finite parameters up 
to linear order in $\bm{k} = (k_x, k_y, k_z)$, and $k_{\pm} = k_x \pm 
ik_y$. For simplicity, let us consider $C = 0$.

For $k_x = k_y = 0$, the  Schrödinger equation $H\psi = \varepsilon_0 \psi$ 
splits into two blocks that can be easily solved for $k_z^{(\pm)} = 
iq_{\pm} = \pm i A_1^{-1}\sqrt{M^2-\varepsilon_0^2}$, where we keep the 
eigenenergy $\varepsilon_0$ as a parameter to be defined by the boundary 
conditions. The four eigenstates associated with $q_{\pm}$ are $\psi_{\pm, 
\nu}(z) = \phi_{\pm, \nu} e^{- q_{\pm} z}$, with

\begin{equation}
    \phi_{\pm,\uparrow} = 
    \begin{pmatrix}
        \pm(\varepsilon_0+M)\\
        \sqrt{\varepsilon_0^2 - M^2}\\
        0 \\ 0
    \end{pmatrix}, \;\;
    \phi_{\pm,\downarrow} = 
    \begin{pmatrix}
        0 \\ 0 \\
        \mp(\varepsilon_0+M)\\
        \sqrt{\varepsilon_0^2 - M^2}\\
    \end{pmatrix}.
\end{equation}

Next, let us use this basis to  obtain the $z=0$ surface states on the 
semi-infinite domain $z\geq 0$. The hard-wall boundary condition 
$(i\mathcal{M}_{\hat{n}} + \mathcal{M}_c)\psi(0) = 0$ (see 
Eq.~\eqref{eq:BC}) for a confinement given by the $M$ terms of 
Eq.~\eqref{eq:HGamma} is set by

\begin{align}
    \mathcal{M}_{\hat{n}} &= 
    \begin{pmatrix}
        0 & -1 & 0 & 0\\
        -1 & 0 & 0 & 0\\
        0 & 0 & 0 & -1\\
        0 & 0 & -1 & 0
    \end{pmatrix},\\
    \mathcal{M}_{c} &= 
    \begin{pmatrix}
        +1 & 0 & 0 & 0\\
        0 & -1 & 0 & 0\\
        0 & 0 & +1 & 0\\
        0 & 0 & 0 & -1
    \end{pmatrix}.
\end{align}
For the $\psi_{\pm,\nu}(z)$ basis,  this boundary condition can only be 
satisfied for $\varepsilon_0 = 0$, thus $q_\pm = \pm |M|/A1$ defines 
surface states with penetration lengths $\ell = A_1/|M|$. Since 
we are looking for semi-infinite surface states at the $z=0$ interface, the 
solutions must vanish at $z \rightarrow \infty$, which selects the $q_+$ 
states as the only physical solutions. The general solution for the $z=0$ 
surface reads

\begin{equation}
    \psi_0(z) = e^{-z/\ell} 
    \left[
    c_1
    \begin{pmatrix}
    -i \\ 1 \\ 0 \\ 0
    \end{pmatrix}
    +
    c_2
    \begin{pmatrix}
    0 \\ 0 \\ i \\ 1
    \end{pmatrix}
    \right],
\end{equation}
where $c_1$ and $c_2$ are arbitrary coefficients.

The same procedure can be applied to  obtain the $z=L$ surface state 
solutions in the domain $z \leq L$. Here the normal to the interface has 
the opposite sign from the $z=0$ solution, hence $\mathcal{M}_{\hat{n}} 
\rightarrow -\mathcal{M}_{\hat{n}}$. The confinement matrix $\mathcal{M}_c$ 
remains the same. In this case, the general solution reads

\begin{equation}
    \psi_L(z) = e^{(z-L)/\ell} 
    \left[
    c_3
    \begin{pmatrix}
    i \\ 1 \\ 0 \\ 0
    \end{pmatrix}
    +
    c_4
    \begin{pmatrix}
    0 \\ 0 \\ -i \\ 1
    \end{pmatrix}
    \right]
\end{equation}
where $c_3$ and $c_4$ are arbitrary coefficients.

Combining $\psi_0(z)$ and $\psi_L(z)$  to form an approximate fourfold 
basis for a thin film on the domain $0 \leq z \leq L$, we project the full 
$H$ from Eq.~\eqref{eq:HGamma} to obtain our effective Hamiltonian 
[Eq.~\eqref{eq:H2D}],

\begin{equation}
    H = \varepsilon_0 +  \hbar v_F(k_x\gamma_{3y}-k_y\gamma_{3x}) + F\gamma_{30} + \Delta\gamma_{10} + B\gamma_{0z},
    \label{eq:Hap}
\end{equation}
where $\hbar v_F = A_2$, $\Delta$ is  an hybridization term that connects 
$c_1$ to $c_4$ and $c_2$ to $c_3$ via the confinement potentials, $F$ is a 
diagonal structural inversion asymmetry (SIA) term that may arise either from an external electric 
field, or due to internal polarization fields as in the interface between 
\BiSe and the metalic lead. The $\mathfrak{su}(2)$ operators $\bm{\sigma}$ 
and $\bm{\tau}$ are introduced to simplify the notation as $\gamma_{ij} = 
\tau_i\otimes\sigma_j$. These act, respectively, on the spin space spanned 
by the $(c_1, c_3)$ up ($\uparrow$) states and $(c_2, c_4)$ down 
($\downarrow$) states, and the top/bottom surfaces spanned by the $(c_1, 
c_2)$ states ($z=0$) and $(c_3, c_4)$ states ($z=L$).


\section{Wilson's mass range}
\label{ap:massrange}

In order to use the Wilson's mass approach to eliminate the doublers, one must choose the value of the Wilson's mass appropriately. A too small value will not eliminate the doublers, while a too large value will deform the low energy spectrum. In this section we establish the approximate lower and upper limits for the Wilson's masses used in the main text.

\begin{figure}[ht!]
  \centering
  \includegraphics[width=0.7\columnwidth,keepaspectratio]{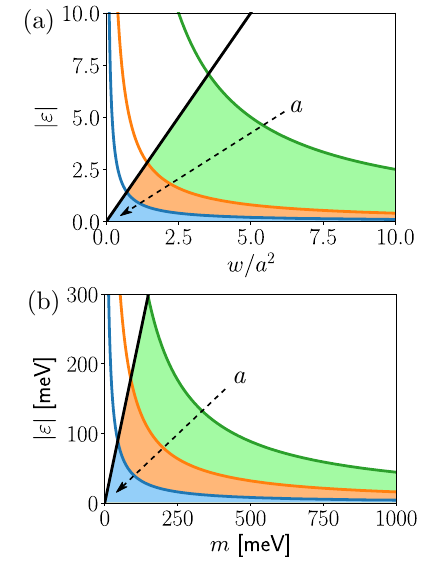}
  \caption{Range of appropriate values of the Wilson mass (shaded areas) for different energies $|\varepsilon|$ and discrete lattice step $a$. (a) For the unidimensional model the range of $a$ is set by Eq.~\eqref{eq:massragenA}, here we set $\hbar v_F = 1$ and $a = \{0.2, 0.5, 1\}$ (following the arrow). (b) For \BiSe the range for the Wilson masses $m = m_B$ or $= m_\Delta$ are set by Eq.~\eqref{eq:massragenB}, with $a = \{3, 5, 10\}$~nm, and $v_F = 479$~nm/ps.}
  \label{Fig:massrange}
\end{figure}

Let us start with the simple unidimensional case defined by $H_\xi + H_W$, where

\begin{align}
    H_\xi &= \hbar v_F \mathcal{M}_\xi k, \\
    H_W   &= w \mathcal{M}_c k^2,
\end{align}
where $\M_\xi$ and $\M_c$ are unitary matrices, $k$ is the momentum along an arbitrary coordinate $\xi$, $v_F$ is the Fermi velocity, and $w$ is Wilson's mass. In a discrete lattice of spacing $a$, the spectrum of $H_\xi + H_W$ has a gap $\Delta\varepsilon = 2w/a^2$ at $|k| = \pi/a$, as shown in Fig.~\ref{Fig1}(a). This energy dispersion approaches the exact linear solution $\varepsilon(k) = \pm \hbar v_F k$ of $H_\xi$ only for $|\varepsilon| \ll \Delta\varepsilon$. This establishes the lower bound $w \gg \frac{1}{2}a^2|\varepsilon|$. Next, the upper bound is obtained by requiring $H_W$ to be a small perturbation to $H_\xi$ is in low energy range. That is, we want $|H_\xi| \gg |H_W|$ for small $\varepsilon$ and $k$. In this limit we can use $|\varepsilon| \approx \hbar v_F k$ to eliminate $k$ from the inequation an obtain $w \ll (\hbar v_F)^2/|\varepsilon|$. Combining these we obtain the range

\begin{equation}
    \frac{1}{2}|\varepsilon| \ll \frac{w}{a^2} \ll \left(\frac{\hbar v_F}{a}\right)^2\frac{1}{|\varepsilon|}.
    \label{eq:massragenA}
\end{equation}
In Fig.~\ref{Fig1}(b) of the main text, the dashed lines correspond to the lower bound rewritten as $|\varepsilon| < 2w/a^2$, while the upper bond does not show up in the range of the figure. In Fig.~\ref{Fig:massrange}(a) we plot this inequation for different values of $a$. 

Now let us discuss the range for $m_B$ or $m_\Delta$ for the \BiSe suraface states from Eqs.~\eqref{eq:H2D} and \eqref{eq:mass}. For simplicity, consider $\varepsilon_0 = F = \Delta = B = 0$ without lack of generality, such that

\begin{align}
    H &= \hbar v_F (k_x \gamma_{3y} - k_y \gamma_{3x}),\\
    H_m &= m \dfrac{a^2}{4} k^2 \gamma_m,
\end{align}
where $(m, \gamma_m) = (m_B, \gamma_{0z})$ for the $B$-type confinement, or $(m, \gamma_m) = (m_\Delta, \gamma_{10})$ for the $\Delta$-type confinement. All $\gamma_{ij}$ matrices are unitary, and $k^2 = k_x^2 + k_y^2$. The factor $a^2/4$ is included so that the mass $m$ have energy units, and labels the gap at $|k_x|$ or $|k_y| = \pi/a$ as $\Delta\varepsilon = 2m$. Equivalently to the previous case, we want to focus on energies $\varepsilon \ll \Delta\varepsilon$, which give us the lower bound $m \gg |\varepsilon|/2$. The upper bound is then obtained requiring $|H| \gg |H_m|$, i.e. $|\varepsilon| \gg |H_m|$. The intensity $|H_m| \approx ma^2k^2/4$, and we can use $|\varepsilon| \approx \hbar v_F |k|$ to replace $k^2$ and obtain $m \ll 4(\hbar v_F/a)^2/|\varepsilon|$. Combining the inequations we get the range

\begin{equation}
    \dfrac{1}{2}|\varepsilon| \ll m \ll \left(\dfrac{2\hbar v_F}{a}\right)^2\dfrac{1}{|\varepsilon|},
    \label{eq:massragenB}
\end{equation}
which only differs from Eq.~\eqref{eq:massragenA} by the factor $2$ in $\hbar v_F$, which is a consequence of the first case beeing a one dimensional model, while the current one is 2D. This range is illustrated in Fig.~\ref{Fig:massrange}(b) for the \BiSe parameters.




\section{Symmetries}
\label{ap:chiral}

Let us discuss the symmetries of our model Hamiltonian $H$ from Eq.~\eqref{eq:H2D} and Eq.~\eqref{eq:Hap} for the \BiSe surface states. For simplicity, we ommit the Wilson mass terms $m_{B/\Delta}$ from Eq.~\eqref{eq:mass} since, regarding the symmetries below, these terms play the same role as the Zeeman field $B$ and surface coupling $\Delta$ terms.

The time-reversal  operator is $\mathcal{T} = e^{-i \frac{\pi}{2} \gamma_{0y}}\mathcal{K} = -i\gamma_{0y}\mathcal{K}$, where $\mathcal{K}$ is the complex conjugation. For the \BiSe Hamiltonian, $[H, \mathcal{T}] = -2B\gamma_{0x}\mathcal{K}$. Therefore only a finite Zeeman term $B$ breaks time-reversal symmetry (TRS) as expected.

A chiral symmetry operator $\Pc$ is a unitary operator that obeys

\begin{equation}
    \{\Pc, H\} = 0, \;\; \Pc\Pc^\dagger = 1, \;\; \Pc^2 = 1,
\end{equation}
where $\{\Pc, H\} = \Pc H + H\Pc$ is the anti-commutator.
The chirality yields a symmetry in the energy spectrum. Given an eigenstate $\phi$ with 
energy $\varepsilon$, there must also exist another eigenstate $\varphi = 
\Pc\phi$ with energy $-\varepsilon$. Aditionally, if 
$H$ admits two chiral symmetries $\Pc$ and $\Pc'$, one can define a 
conserved chiral charge $\Q = \Pc\Pc'$ such that it commutes with $H$, i.e. $[\Q, H] = \Q H-H\Q = 0$.
Therefore, the eigenvalues $q$ of $\Q$ can be used to classify the eigenstates and block-diaginalize $H$.

Here we find four candidates for chiral operators: $\Pc_{0z} = \gamma_{0z}$, $\Pc_{10} = \gamma_{10}$, 
$\Pc_{20} = \gamma_{20}$, and $\Pc_{3z} = \gamma_{3z}$. These obey

\begin{align}
 \{\Pc_{0z}, H-\varepsilon_0\} &= 2B + 2\Delta\gamma_{1z} + 2F\gamma_{3z},\\
 \{\Pc_{10}, H-\varepsilon_0\} &= 2B\gamma_{1z} + 2\Delta,\\
 \{\Pc_{20}, H-\varepsilon_0\} &= 2B\gamma_{2z},\\
 \{\Pc_{3z}, H-\varepsilon_0\} &= 2B\gamma_{30} + 2F\gamma_{0z},
\end{align}
where the rigid shift $\varepsilon_0$ only affects the energy symmetry point. Notice that a finite $B$ (or $m_B$) breaks all chiralities above. Which of these operators represent chiral symmetries of $H$ depend on which terms ($B$, $F$, and $\Delta$) are zero. Additionally, these chiral operators transform by TRS as

\begin{align}
 \{\Pc_{0z}, \mathcal{T}\} &= 0,\\
 [\Pc_{10}, \mathcal{T}] &= 0,\\
 \{\Pc_{20}, \mathcal{T}\} &= 0,\\
 \{\Pc_{3z}, \mathcal{T}\} &= 0.
\end{align}

\subsection{Chiralities}

In the main text, Fig.~\ref{Fig2}(a) show the energy spectrum for a circular quantum dot with $\Delta$-type hard-wall confinement and $B=F=0$. In this case $\Pc_{20}$ and $\Pc_{3z}$ are chiral symmetries of $H$. Additionally, TRS and the total angular momentum $J_z$ are preserved. This allow us to define a chiral charge $\Q_{1z} = -i \Pc_{20}\Pc_{3z} = \gamma_{1z}$, whose eigenvalues are $q_{1z} = \pm 1$. Both chiralities commute with $J_z$, $[\Pc_{20}, J_z] = [\Pc_{3z}, J_z] = 0$, and anti-commute with $\Q_{1z}$, $\{\Pc_{20}, \Q_{1z}\} = \{\Pc_{3z}, \Q_{1z}\} = 0$. Consequently, a state $\ket{j_z, q_{1z}}$ with energy $\varepsilon_{j_z, q_{1z}}$ has a chiral partner $\ket{j_z, -q_{1z}}$ with energy $\varepsilon_{j_z, -q_{1z}} = -\varepsilon_{j_z, q_{1z}}$, the same angular momentum $j_z$ and opposite chirality $-q_{1z}$. The TRS anti-commutes with both $J_z$ and $\Q_{1z}$, i.e. $\{\mathcal{T}, J_z\} = \{\mathcal{T}, \Q_{1z}\} = 0$. Therefore, the Kramer partner of $\ket{j_z, q_{1z}}$ is the state $\ket{-j_z, -q_{1z}}$ with energy $\varepsilon_{-j_z, -q_{1z}} = \varepsilon_{j_z, q_{1z}}$. These parters are illustrated in Fig.~\ref{Fig:dotchiral}(a).

In contrast with the case above, the energy spectrum in Fig.~\ref{Fig2}(b) refers to a circular quantum dot with a $B$-type hard-wall confinement. Here, the finite Wilson mass $m_B$ breaks both TRS and all chiral symmetries above. However, the product $\Pt_j = \Pc_j\mathcal{T}$ of each $\Pc_j$ above with the time-reversal operator $\mathcal{T}$ is preserved. We refer to these as ``time-reversal chiralities'', which obey

\begin{equation}
    \Pt_j{\Pt_j}^\dagger = 1, \;\; {\Pt_j}^2 = \mp 1,
\end{equation}
where the sign $\mp$ on the second expression refer to the cases where $[\Pc_j, \mathcal{T}] = 0$, or $\{\Pc_j, \mathcal{T}\} = 0$, respectively. Notice that if $\Pc_j$ commutes (anti-commutes) with $\mathcal{T}$, then $\Pt_j$ anti-commutes (commutes) with $\mathcal{T}$. From the four chiral candidates $\Pc_j$ above, only $\Pc_{10}$ commutes with $\mathcal{T}$, yielding ${\Pt_{10}}^2 = -1$, which does not fall into the chiral classification \cite{PhysRevB.55.1142, bernard2002classification, Schnyder2008PRBClassification}. The three other $\Pt_j$ are candidates for chiral operators. For completeness, we write the anti-commutation of all four $\Pt_j$ with $H$ as

\begin{align}
 \{\Pt_{0z}, H - \varepsilon_0\} &= (-2\Delta\gamma_{1x} - 2F\gamma_{3x})\mathcal{K},\\
 \{\Pt_{10}, H - \varepsilon_0\} &= (-2i\Delta\gamma_{0y})\mathcal{K},\\
 \{\Pt_{20}, H - \varepsilon_0\} &= 0,\\
 \{\Pt_{3z}, H - \varepsilon_0\} &= (-2F\gamma_{0x})\mathcal{K}.
\end{align}
Interestingly, these anti-commutators do not depend upon $B$, and $\Pt_{20}$ is always a chiral operator, independent of the model parameters.

\begin{figure}[ht!]
  \centering
  \includegraphics[width=\columnwidth,keepaspectratio]{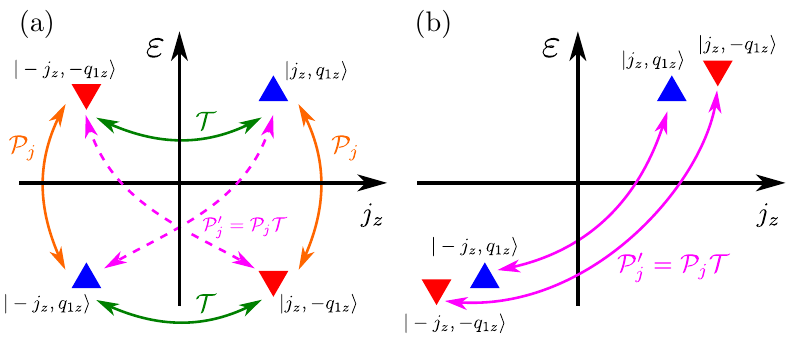}
  \caption{Energy-angular momentum diagram of the eigenstates of the cylindrical dot. The up (down) triangles label the chiral charges $q_{1z} = \pm 1$ as in Fig.~\ref{Fig2} of the main text. (a) For the $\Delta$-type confinement at least one chiral symmetry $\Pc_j$ and TRS $\mathcal{T}$ are present. As indicanted in the diagram, the chiral operators $\Pc_j$ transform the states fliping their chiral charge and energy, but preserve the total angular momentum $j_z$, while TRS connects the Kramer partners. (b) For the $B$-type confinement both TRS and the $\Pc_j$ symmetries are broken, but their product $\Pt_j = \Pc_j \mathcal{T}$ is preserved as a ``time-reversal chiral'' symmetry that connects states preserving their charge, but flipping both $j_z$ and energy. In (a) the $\Pt_j$ is also present, since its components are preserved symmetries as well.
  }
  \label{Fig:dotchiral}
\end{figure}

For the case of Fig.~\ref{Fig2}(a) we have $F = 0$, $m_\Delta = 0$, $\Delta \neq 0$ and $m_B \neq 0$. Therefore, the chiral operators are $\Pt_{20}$ and $\Pt_{3z}$. These combine to define the same chiral charge as before, i.e. $\Q_{1z} = -i\Pt_{20}\Pt_{3z} = \gamma_{1z}$. However, now they anti-commute with $J_z$, $\{\Pt_{20}, J_z\} = \{\Pt_{3z}, J_z\} = 0$, and commute with $\Q_{1z}$, $[\Pt_{20}, \Q_{1z}] = [\Pt_{3z}, \Q_{1z}] = 0$. Consequently, the state $\ket{j_z, q_{1z}}$ with energy $\varepsilon_{j_z,q_{1z}}$ has a chiral partner $\ket{-j_z, q_{1z}}$ with energy $\varepsilon_{-j_z, q_{1z}} = -\varepsilon_{j_z,q_{1z}}$ and the same charge, as illustrated in Fig.~\ref{Fig:dotchiral}(b). Since TRS is broken by $m_B$, there's no Kramer partners in this case.

\subsection{Surface localization}
\label{S:localization}

\begin{figure}[ht!]
    \centering
    \includegraphics[width=\columnwidth,keepaspectratio]{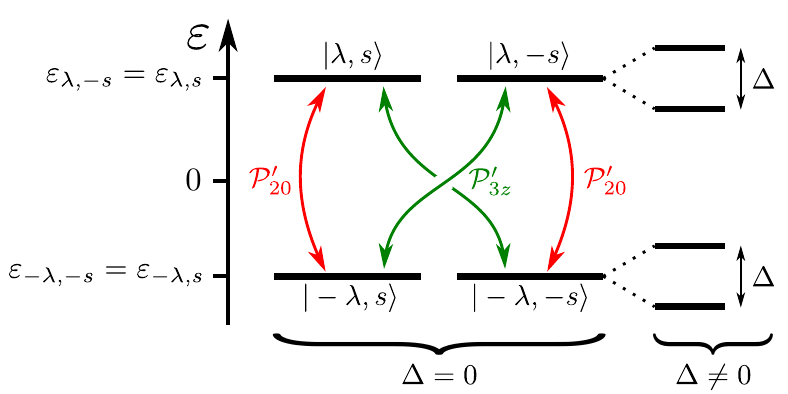}
    \caption{Energy diagram of the eigenstates $\ket{\lambda, s}$, where $\lambda = \pm 1$ labels the energy sign and $s$ are the eigenvalues of the surface operator $S = \gamma_{30}$, with $s = +1 (-1)$ for the top (bottom) surface. On the left, for $\Delta = 0$, the chirality $\Pt_{20}$ transform the states flipping the sign of $\lambda$ in a fixed surface $s$, while $\Pt_{3z}$ flips both $\lambda$ and the surface $s$. On the right a finite $\Delta$ hybridizes the surfaces, thus splitting the degenerate levels.
    }
    \label{Fig:mbchiral}
\end{figure}

Apart from $J_z$, all other symmetries discussed above for the cylindrical quantum dot holds for the transport properties discussion that follows Fig.~\ref{Fig4} in the main text. Additionally, since there $\Delta$ is either zero or small, it is interesting to discuss the localization of the surface states. For such, let us define the surface operator $S = \gamma_{30}$, whose eigenvalues $s = \pm 1$ refer to the top and bottom surfaces. Evidently, $s$ is a good quantum number only if $\Delta$ (and $m_\Delta$) are zero, otherwise

\begin{equation}
    [H, S] = 2 i \Delta \gamma_{20}.
\end{equation}

Let us consider the $B$-type confinement and $[H,S]=0$. Since TRS is broken by $m_B$, the relevant chiralities are $\Pt_{20}$ and $\Pt_{3z}$, thus yielding the conserved charge $\Q_{1z}$, i.e. $[H, \Q_{1z}] = 0$. However, since $\{S, \Q_{1z}\} = 0$, there are no common basis between $H$, $\Q_{1z}$, and $S$. In the previous discussion regarding the chiral symmetries on the cylindrical quantum dot we have used a common basis between $H$, $\Q_{1z}$ and $J_z$. Instead, hereafter we shall consider a common basis between $H$ and $S$, which we label as $\ket{\lambda, s}$, where $s = \pm 1$ is the surface eigenvalue, and $\lambda = \pm 1$ will distinguish the chiral partners. Interestingly, we find $[\Pt_{20}, S] = 0$ and $\{\Pt_{3z}, S\} =0$, which tell us that each $\ket{\lambda, s}$ have two chiral partners: (i) $\ket{-\lambda, s} = \Pt_{20}\ket{\lambda, s}$ in the same surface, but with energy $\varepsilon_{-\lambda, s} = -\varepsilon_{\lambda,s}$; and (ii) $\ket{-\lambda, -s} = \Pt_{3z}\ket{\lambda,s}$ in the opposite surface and energy $\varepsilon_{-\lambda,-s} = -\varepsilon_{\lambda,s}$. Moreover, combining these two symmetries, the relation $\{S, \Q_{1z}\} = 0$ give us a fourth state, which we dub as a ``charge'' partner given by $\ket{\lambda, -s} = \Q_{1z}\ket{\lambda, s}$ in opposite surface and degenerate in energy, i.e. $\varepsilon_{\lambda, -s} = +\varepsilon_{\lambda,s}$. These transformations are illustrated in Fig.~\ref{Fig:mbchiral}. For a small, but finite $\Delta$, the degenerate eigenstates hybridize and split as seen in Fig.~\ref{Fig4}(b) in the main text. 

}

\bibliography{refs}

\begin{thebibliography}{64}%
\makeatletter
\providecommand \@ifxundefined [1]{%
 \@ifx{#1\undefined}
}%
\providecommand \@ifnum [1]{%
 \ifnum #1\expandafter \@firstoftwo
 \else \expandafter \@secondoftwo
 \fi
}%
\providecommand \@ifx [1]{%
 \ifx #1\expandafter \@firstoftwo
 \else \expandafter \@secondoftwo
 \fi
}%
\providecommand \natexlab [1]{#1}%
\providecommand \enquote  [1]{``#1''}%
\providecommand \bibnamefont  [1]{#1}%
\providecommand \bibfnamefont [1]{#1}%
\providecommand \citenamefont [1]{#1}%
\providecommand \href@noop [0]{\@secondoftwo}%
\providecommand \href [0]{\begingroup \@sanitize@url \@href}%
\providecommand \@href[1]{\@@startlink{#1}\@@href}%
\providecommand \@@href[1]{\endgroup#1\@@endlink}%
\providecommand \@sanitize@url [0]{\catcode `\\12\catcode `\$12\catcode
  `\&12\catcode `\#12\catcode `\^12\catcode `\_12\catcode `\%12\relax}%
\providecommand \@@startlink[1]{}%
\providecommand \@@endlink[0]{}%
\providecommand \url  [0]{\begingroup\@sanitize@url \@url }%
\providecommand \@url [1]{\endgroup\@href {#1}{\urlprefix }}%
\providecommand \urlprefix  [0]{URL }%
\providecommand \Eprint [0]{\href }%
\providecommand \doibase [0]{http://dx.doi.org/}%
\providecommand \selectlanguage [0]{\@gobble}%
\providecommand \bibinfo  [0]{\@secondoftwo}%
\providecommand \bibfield  [0]{\@secondoftwo}%
\providecommand \translation [1]{[#1]}%
\providecommand \BibitemOpen [0]{}%
\providecommand \bibitemStop [0]{}%
\providecommand \bibitemNoStop [0]{.\EOS\space}%
\providecommand \EOS [0]{\spacefactor3000\relax}%
\providecommand \BibitemShut  [1]{\csname bibitem#1\endcsname}%
\let\auto@bib@innerbib\@empty
\bibitem [{\citenamefont {Bernevig}\ and\ \citenamefont
  {Zhang}(2006)}]{PhysRevLett.96.106802}%
  \BibitemOpen
  \bibfield  {author} {\bibinfo {author} {\bibfnamefont {B.~A.}\ \bibnamefont
  {Bernevig}}\ and\ \bibinfo {author} {\bibfnamefont {S.-C.}\ \bibnamefont
  {Zhang}},\ }\href {\doibase 10.1103/PhysRevLett.96.106802} {\bibfield
  {journal} {\bibinfo  {journal} {Phys. Rev. Lett.}\ }\textbf {\bibinfo
  {volume} {96}},\ \bibinfo {pages} {106802} (\bibinfo {year}
  {2006})}\BibitemShut {NoStop}%
\bibitem [{\citenamefont {K{\"o}nig}\ \emph {et~al.}(2007)\citenamefont
  {K{\"o}nig}, \citenamefont {Wiedmann}, \citenamefont {Br{\"u}ne},
  \citenamefont {Roth}, \citenamefont {Buhmann}, \citenamefont {Molenkamp},
  \citenamefont {Qi},\ and\ \citenamefont {Zhang}}]{Konig766}%
  \BibitemOpen
  \bibfield  {author} {\bibinfo {author} {\bibfnamefont {M.}~\bibnamefont
  {K{\"o}nig}}, \bibinfo {author} {\bibfnamefont {S.}~\bibnamefont {Wiedmann}},
  \bibinfo {author} {\bibfnamefont {C.}~\bibnamefont {Br{\"u}ne}}, \bibinfo
  {author} {\bibfnamefont {A.}~\bibnamefont {Roth}}, \bibinfo {author}
  {\bibfnamefont {H.}~\bibnamefont {Buhmann}}, \bibinfo {author} {\bibfnamefont
  {L.~W.}\ \bibnamefont {Molenkamp}}, \bibinfo {author} {\bibfnamefont {X.-L.}\
  \bibnamefont {Qi}}, \ and\ \bibinfo {author} {\bibfnamefont {S.-C.}\
  \bibnamefont {Zhang}},\ }\href {\doibase 10.1126/science.1148047} {\bibfield
  {journal} {\bibinfo  {journal} {Science}\ }\textbf {\bibinfo {volume}
  {318}},\ \bibinfo {pages} {766} (\bibinfo {year} {2007})}\BibitemShut
  {NoStop}%
\bibitem [{\citenamefont {Hasan}\ and\ \citenamefont
  {Kane}(2010)}]{RevModPhys.82.3045}%
  \BibitemOpen
  \bibfield  {author} {\bibinfo {author} {\bibfnamefont {M.~Z.}\ \bibnamefont
  {Hasan}}\ and\ \bibinfo {author} {\bibfnamefont {C.~L.}\ \bibnamefont
  {Kane}},\ }\href {\doibase 10.1103/RevModPhys.82.3045} {\bibfield  {journal}
  {\bibinfo  {journal} {Rev. Mod. Phys.}\ }\textbf {\bibinfo {volume} {82}},\
  \bibinfo {pages} {3045} (\bibinfo {year} {2010})}\BibitemShut {NoStop}%
\bibitem [{\citenamefont {Qi}\ and\ \citenamefont
  {Zhang}(2011)}]{RevModPhys.83.1057}%
  \BibitemOpen
  \bibfield  {author} {\bibinfo {author} {\bibfnamefont {X.-L.}\ \bibnamefont
  {Qi}}\ and\ \bibinfo {author} {\bibfnamefont {S.-C.}\ \bibnamefont {Zhang}},\
  }\href {\doibase 10.1103/RevModPhys.83.1057} {\bibfield  {journal} {\bibinfo
  {journal} {Rev. Mod. Phys.}\ }\textbf {\bibinfo {volume} {83}},\ \bibinfo
  {pages} {1057} (\bibinfo {year} {2011})}\BibitemShut {NoStop}%
\bibitem [{\citenamefont {G\"otte}\ \emph {et~al.}(2014)\citenamefont
  {G\"otte}, \citenamefont {Paananen}, \citenamefont {Reiss},\ and\
  \citenamefont {Dahm}}]{PhysRevApplied.2.054010}%
  \BibitemOpen
  \bibfield  {author} {\bibinfo {author} {\bibfnamefont {M.}~\bibnamefont
  {G\"otte}}, \bibinfo {author} {\bibfnamefont {T.}~\bibnamefont {Paananen}},
  \bibinfo {author} {\bibfnamefont {G.}~\bibnamefont {Reiss}}, \ and\ \bibinfo
  {author} {\bibfnamefont {T.}~\bibnamefont {Dahm}},\ }\href {\doibase
  10.1103/PhysRevApplied.2.054010} {\bibfield  {journal} {\bibinfo  {journal}
  {Phys. Rev. Appl.}\ }\textbf {\bibinfo {volume} {2}},\ \bibinfo {pages}
  {054010} (\bibinfo {year} {2014})}\BibitemShut {NoStop}%
\bibitem [{\citenamefont {G{\"o}tte}\ \emph {et~al.}(2016)\citenamefont
  {G{\"o}tte}, \citenamefont {Joppe},\ and\ \citenamefont {Dahm}}]{Thomas2016}%
  \BibitemOpen
  \bibfield  {author} {\bibinfo {author} {\bibfnamefont {M.}~\bibnamefont
  {G{\"o}tte}}, \bibinfo {author} {\bibfnamefont {M.}~\bibnamefont {Joppe}}, \
  and\ \bibinfo {author} {\bibfnamefont {T.}~\bibnamefont {Dahm}},\ }\href
  {\doibase 10.1038/srep36070} {\bibfield  {journal} {\bibinfo  {journal} {Sci.
  Rep.}\ }\textbf {\bibinfo {volume} {6}},\ \bibinfo {pages} {36070} (\bibinfo
  {year} {2016})}\BibitemShut {NoStop}%
\bibitem [{\citenamefont {Paudel}\ and\ \citenamefont
  {Leuenberger}(2013)}]{PhysRevB.88.085316}%
  \BibitemOpen
  \bibfield  {author} {\bibinfo {author} {\bibfnamefont {H.~P.}\ \bibnamefont
  {Paudel}}\ and\ \bibinfo {author} {\bibfnamefont {M.~N.}\ \bibnamefont
  {Leuenberger}},\ }\href {\doibase 10.1103/PhysRevB.88.085316} {\bibfield
  {journal} {\bibinfo  {journal} {Phys. Rev. B}\ }\textbf {\bibinfo {volume}
  {88}},\ \bibinfo {pages} {085316} (\bibinfo {year} {2013})}\BibitemShut
  {NoStop}%
\bibitem [{\citenamefont {Yeats}\ \emph {et~al.}(2015)\citenamefont {Yeats},
  \citenamefont {Pan}, \citenamefont {Richardella}, \citenamefont {Mintun},
  \citenamefont {Samarth},\ and\ \citenamefont {Awschalom}}]{Yeatse1500640}%
  \BibitemOpen
  \bibfield  {author} {\bibinfo {author} {\bibfnamefont {A.~L.}\ \bibnamefont
  {Yeats}}, \bibinfo {author} {\bibfnamefont {Y.}~\bibnamefont {Pan}}, \bibinfo
  {author} {\bibfnamefont {A.}~\bibnamefont {Richardella}}, \bibinfo {author}
  {\bibfnamefont {P.~J.}\ \bibnamefont {Mintun}}, \bibinfo {author}
  {\bibfnamefont {N.}~\bibnamefont {Samarth}}, \ and\ \bibinfo {author}
  {\bibfnamefont {D.~D.}\ \bibnamefont {Awschalom}},\ }\href {\doibase
  10.1126/sciadv.1500640} {\bibfield  {journal} {\bibinfo  {journal} {Sci.
  Adv.}\ }\textbf {\bibinfo {volume} {1}},\ \bibinfo {pages} {e1500640}
  (\bibinfo {year} {2015})}\BibitemShut {NoStop}%
\bibitem [{\citenamefont {Kim}\ \emph {et~al.}(2011)\citenamefont {Kim},
  \citenamefont {Brahlek}, \citenamefont {Bansal}, \citenamefont {Edrey},
  \citenamefont {Kapilevich}, \citenamefont {Iida}, \citenamefont {Tanimura},
  \citenamefont {Horibe}, \citenamefont {Cheong},\ and\ \citenamefont
  {Oh}}]{PhysRevB.84.073109}%
  \BibitemOpen
  \bibfield  {author} {\bibinfo {author} {\bibfnamefont {Y.~S.}\ \bibnamefont
  {Kim}}, \bibinfo {author} {\bibfnamefont {M.}~\bibnamefont {Brahlek}},
  \bibinfo {author} {\bibfnamefont {N.}~\bibnamefont {Bansal}}, \bibinfo
  {author} {\bibfnamefont {E.}~\bibnamefont {Edrey}}, \bibinfo {author}
  {\bibfnamefont {G.~A.}\ \bibnamefont {Kapilevich}}, \bibinfo {author}
  {\bibfnamefont {K.}~\bibnamefont {Iida}}, \bibinfo {author} {\bibfnamefont
  {M.}~\bibnamefont {Tanimura}}, \bibinfo {author} {\bibfnamefont
  {Y.}~\bibnamefont {Horibe}}, \bibinfo {author} {\bibfnamefont {S.-W.}\
  \bibnamefont {Cheong}}, \ and\ \bibinfo {author} {\bibfnamefont
  {S.}~\bibnamefont {Oh}},\ }\href {\doibase 10.1103/PhysRevB.84.073109}
  {\bibfield  {journal} {\bibinfo  {journal} {Phys. Rev. B}\ }\textbf {\bibinfo
  {volume} {84}},\ \bibinfo {pages} {073109} (\bibinfo {year}
  {2011})}\BibitemShut {NoStop}%
\bibitem [{\citenamefont {Mishra}\ \emph {et~al.}(1997)\citenamefont {Mishra},
  \citenamefont {Satpathy},\ and\ \citenamefont {Jepsen}}]{0953-8984-9-2-014}%
  \BibitemOpen
  \bibfield  {author} {\bibinfo {author} {\bibfnamefont {S.~K.}\ \bibnamefont
  {Mishra}}, \bibinfo {author} {\bibfnamefont {S.}~\bibnamefont {Satpathy}}, \
  and\ \bibinfo {author} {\bibfnamefont {O.}~\bibnamefont {Jepsen}},\ }\href
  {\doibase 10.1088/0953-8984/9/2/014} {\bibfield  {journal} {\bibinfo
  {journal} {J. Phys.: Condens. Matter}\ }\textbf {\bibinfo {volume} {9}},\
  \bibinfo {pages} {461} (\bibinfo {year} {1997})}\BibitemShut {NoStop}%
\bibitem [{\citenamefont {Hao}\ \emph {et~al.}(2011)\citenamefont {Hao},
  \citenamefont {Thalmeier},\ and\ \citenamefont {Lee}}]{hao2011topological}%
  \BibitemOpen
  \bibfield  {author} {\bibinfo {author} {\bibfnamefont {L.}~\bibnamefont
  {Hao}}, \bibinfo {author} {\bibfnamefont {P.}~\bibnamefont {Thalmeier}}, \
  and\ \bibinfo {author} {\bibfnamefont {T.~K.}\ \bibnamefont {Lee}},\ }\href
  {\doibase 10.1103/PhysRevB.84.235303} {\bibfield  {journal} {\bibinfo
  {journal} {Phys. Rev. B}\ }\textbf {\bibinfo {volume} {84}},\ \bibinfo
  {pages} {235303} (\bibinfo {year} {2011})}\BibitemShut {NoStop}%
\bibitem [{\citenamefont {DeGrand}(2007)}]{DeGrand2007}%
  \BibitemOpen
  \bibfield  {author} {\bibinfo {author} {\bibfnamefont {T.}~\bibnamefont
  {DeGrand}},\ }\href {\doibase 10.1140/epjst/e2007-00375-4} {\bibfield
  {journal} {\bibinfo  {journal} {Eur. Phys. J. Spec. Top.}\ }\textbf {\bibinfo
  {volume} {152}},\ \bibinfo {pages} {1} (\bibinfo {year} {2007})}\BibitemShut
  {NoStop}%
\bibitem [{\citenamefont {Chandrasekharan}\ and\ \citenamefont
  {Wiese}(2004)}]{chandrasekharan2004introduction}%
  \BibitemOpen
  \bibfield  {author} {\bibinfo {author} {\bibfnamefont {S.}~\bibnamefont
  {Chandrasekharan}}\ and\ \bibinfo {author} {\bibfnamefont {U.-J.}\
  \bibnamefont {Wiese}},\ }\href {\doibase 10.1016/j.ppnp.2004.05.003}
  {\bibfield  {journal} {\bibinfo  {journal} {Prog. Part. Nucl. Phys.}\
  }\textbf {\bibinfo {volume} {53}},\ \bibinfo {pages} {373} (\bibinfo {year}
  {2004})}\BibitemShut {NoStop}%
\bibitem [{\citenamefont {Susskind}(1977)}]{Susskind1977Staggered}%
  \BibitemOpen
  \bibfield  {author} {\bibinfo {author} {\bibfnamefont {L.}~\bibnamefont
  {Susskind}},\ }\href {\doibase 10.1103/PhysRevD.16.3031} {\bibfield
  {journal} {\bibinfo  {journal} {Phys. Rev. D}\ }\textbf {\bibinfo {volume}
  {16}},\ \bibinfo {pages} {3031} (\bibinfo {year} {1977})}\BibitemShut
  {NoStop}%
\bibitem [{\citenamefont {Stacey}(1982)}]{stacey82}%
  \BibitemOpen
  \bibfield  {author} {\bibinfo {author} {\bibfnamefont {R.}~\bibnamefont
  {Stacey}},\ }\href {\doibase 10.1103/PhysRevD.26.468} {\bibfield  {journal}
  {\bibinfo  {journal} {Phys. Rev. D}\ }\textbf {\bibinfo {volume} {26}},\
  \bibinfo {pages} {468} (\bibinfo {year} {1982})}\BibitemShut {NoStop}%
\bibitem [{\citenamefont {Tworzyd\l{}o}\ \emph {et~al.}(2008)\citenamefont
  {Tworzyd\l{}o}, \citenamefont {Groth},\ and\ \citenamefont
  {Beenakker}}]{finitedifference2008}%
  \BibitemOpen
  \bibfield  {author} {\bibinfo {author} {\bibfnamefont {J.}~\bibnamefont
  {Tworzyd\l{}o}}, \bibinfo {author} {\bibfnamefont {C.~W.}\ \bibnamefont
  {Groth}}, \ and\ \bibinfo {author} {\bibfnamefont {C.~W.~J.}\ \bibnamefont
  {Beenakker}},\ }\href {\doibase 10.1103/PhysRevB.78.235438} {\bibfield
  {journal} {\bibinfo  {journal} {Phys. Rev. B}\ }\textbf {\bibinfo {volume}
  {78}},\ \bibinfo {pages} {235438} (\bibinfo {year} {2008})}\BibitemShut
  {NoStop}%
\bibitem [{\citenamefont {Hern{\'a}ndez}\ and\ \citenamefont
  {Lewenkopf}(2012)}]{hernandez2012finite}%
  \BibitemOpen
  \bibfield  {author} {\bibinfo {author} {\bibfnamefont {A.~R.}\ \bibnamefont
  {Hern{\'a}ndez}}\ and\ \bibinfo {author} {\bibfnamefont {C.~H.}\ \bibnamefont
  {Lewenkopf}},\ }\href {\doibase 10.1103/PhysRevB.86.155439} {\bibfield
  {journal} {\bibinfo  {journal} {Phys. Rev. B}\ }\textbf {\bibinfo {volume}
  {86}},\ \bibinfo {pages} {155439} (\bibinfo {year} {2012})}\BibitemShut
  {NoStop}%
\bibitem [{\citenamefont {Wilson}(1974)}]{wilson1974confinement}%
  \BibitemOpen
  \bibfield  {author} {\bibinfo {author} {\bibfnamefont {K.~G.}\ \bibnamefont
  {Wilson}},\ }\href {\doibase 10.1103/PhysRevD.10.2445} {\bibfield  {journal}
  {\bibinfo  {journal} {Phys. Rev. D}\ }\textbf {\bibinfo {volume} {10}},\
  \bibinfo {pages} {2445} (\bibinfo {year} {1974})}\BibitemShut {NoStop}%
\bibitem [{\citenamefont {Kogut}\ and\ \citenamefont
  {Susskind}(1975)}]{Kogut1975}%
  \BibitemOpen
  \bibfield  {author} {\bibinfo {author} {\bibfnamefont {J.~B.}\ \bibnamefont
  {Kogut}}\ and\ \bibinfo {author} {\bibfnamefont {L.}~\bibnamefont
  {Susskind}},\ }\href {\doibase 10.1103/PhysRevD.11.395} {\bibfield  {journal}
  {\bibinfo  {journal} {Phys. Rev. D}\ }\textbf {\bibinfo {volume} {11}},\
  \bibinfo {pages} {395} (\bibinfo {year} {1975})}\BibitemShut {NoStop}%
\bibitem [{\citenamefont {Bermudez}\ \emph {et~al.}(2010)\citenamefont
  {Bermudez}, \citenamefont {Mazza}, \citenamefont {Rizzi}, \citenamefont
  {Goldman}, \citenamefont {Lewenstein},\ and\ \citenamefont
  {Martin-Delgado}}]{wilsonfermionsoptical}%
  \BibitemOpen
  \bibfield  {author} {\bibinfo {author} {\bibfnamefont {A.}~\bibnamefont
  {Bermudez}}, \bibinfo {author} {\bibfnamefont {L.}~\bibnamefont {Mazza}},
  \bibinfo {author} {\bibfnamefont {M.}~\bibnamefont {Rizzi}}, \bibinfo
  {author} {\bibfnamefont {N.}~\bibnamefont {Goldman}}, \bibinfo {author}
  {\bibfnamefont {M.}~\bibnamefont {Lewenstein}}, \ and\ \bibinfo {author}
  {\bibfnamefont {M.~A.}\ \bibnamefont {Martin-Delgado}},\ }\href {\doibase
  10.1103/PhysRevLett.105.190404} {\bibfield  {journal} {\bibinfo  {journal}
  {Phys. Rev. Lett.}\ }\textbf {\bibinfo {volume} {105}},\ \bibinfo {pages}
  {190404} (\bibinfo {year} {2010})}\BibitemShut {NoStop}%
\bibitem [{\citenamefont {Zhou}\ \emph {et~al.}(2017)\citenamefont {Zhou},
  \citenamefont {Jiang}, \citenamefont {Xie},\ and\ \citenamefont
  {Sun}}]{zhou2016LatticeModel}%
  \BibitemOpen
  \bibfield  {author} {\bibinfo {author} {\bibfnamefont {Y.-F.}\ \bibnamefont
  {Zhou}}, \bibinfo {author} {\bibfnamefont {H.}~\bibnamefont {Jiang}},
  \bibinfo {author} {\bibfnamefont {X.~C.}\ \bibnamefont {Xie}}, \ and\
  \bibinfo {author} {\bibfnamefont {Q.-F.}\ \bibnamefont {Sun}},\ }\href
  {\doibase 10.1103/PhysRevB.95.245137} {\bibfield  {journal} {\bibinfo
  {journal} {Phys. Rev. B}\ }\textbf {\bibinfo {volume} {95}},\ \bibinfo
  {pages} {245137} (\bibinfo {year} {2017})}\BibitemShut {NoStop}%
\bibitem [{\citenamefont {Svetitsky}\ \emph {et~al.}(1980)\citenamefont
  {Svetitsky}, \citenamefont {Drell}, \citenamefont {Quinn},\ and\
  \citenamefont {Weinstein}}]{Svetitsky1980SLAC}%
  \BibitemOpen
  \bibfield  {author} {\bibinfo {author} {\bibfnamefont {B.}~\bibnamefont
  {Svetitsky}}, \bibinfo {author} {\bibfnamefont {S.~D.}\ \bibnamefont
  {Drell}}, \bibinfo {author} {\bibfnamefont {H.~R.}\ \bibnamefont {Quinn}}, \
  and\ \bibinfo {author} {\bibfnamefont {M.}~\bibnamefont {Weinstein}},\ }\href
  {\doibase 10.1103/PhysRevD.22.490} {\bibfield  {journal} {\bibinfo  {journal}
  {Phys. Rev. D}\ }\textbf {\bibinfo {volume} {22}},\ \bibinfo {pages} {490}
  (\bibinfo {year} {1980})}\BibitemShut {NoStop}%
\bibitem [{\citenamefont {Quinn}\ and\ \citenamefont
  {Weinstein}(1986)}]{Quinn1986SLAC}%
  \BibitemOpen
  \bibfield  {author} {\bibinfo {author} {\bibfnamefont {H.~R.}\ \bibnamefont
  {Quinn}}\ and\ \bibinfo {author} {\bibfnamefont {M.}~\bibnamefont
  {Weinstein}},\ }\href {\doibase 10.1103/PhysRevLett.57.2617} {\bibfield
  {journal} {\bibinfo  {journal} {Phys. Rev. Lett.}\ }\textbf {\bibinfo
  {volume} {57}},\ \bibinfo {pages} {2617} (\bibinfo {year}
  {1986})}\BibitemShut {NoStop}%
\bibitem [{SLA()}]{SLAC12}%
  \BibitemOpen
  \href@noop {} {}\bibinfo {note} {J.~P.~Costella,~
  \href{https://arxiv.org/abs/hep-lat/0207008}{arXiv:hep-lat/0207008}.~
  \href{https://arxiv.org/abs/hep-lat/0207015}{arXiv:hep-lat/0207015}.}\BibitemShut
  {Stop}%
\bibitem [{\citenamefont {Kaplan}(1992)}]{kaplan1992method}%
  \BibitemOpen
  \bibfield  {author} {\bibinfo {author} {\bibfnamefont {D.~B.}\ \bibnamefont
  {Kaplan}},\ }\href {\doibase 10.1016/0370-2693(92)91112-M} {\bibfield
  {journal} {\bibinfo  {journal} {Phys. Lett. B}\ }\textbf {\bibinfo {volume}
  {288}},\ \bibinfo {pages} {342} (\bibinfo {year} {1992})}\BibitemShut
  {NoStop}%
\bibitem [{\citenamefont {Kaplan}\ and\ \citenamefont
  {Sun}(2012)}]{Kaplan2012Spacetime}%
  \BibitemOpen
  \bibfield  {author} {\bibinfo {author} {\bibfnamefont {D.~B.}\ \bibnamefont
  {Kaplan}}\ and\ \bibinfo {author} {\bibfnamefont {S.}~\bibnamefont {Sun}},\
  }\href {\doibase 10.1103/PhysRevLett.108.181807} {\bibfield  {journal}
  {\bibinfo  {journal} {Phys. Rev. Lett.}\ }\textbf {\bibinfo {volume} {108}},\
  \bibinfo {pages} {181807} (\bibinfo {year} {2012})}\BibitemShut {NoStop}%
\bibitem [{cre()}]{creutz1994Talk}%
  \BibitemOpen
  \href@noop {} {}\bibinfo {note} {M. Creutz,
  \href{https://arxiv.org/abs/hep-lat/9410008}{arXiv:hep-lat/9410008}.
  \href{https://doi.org/10.1016/0920-5632(95)00187-E}{Nucl. Phys. B 42, 56
  (1995)}.}\BibitemShut {Stop}%
\bibitem [{\citenamefont {Creutz}\ and\ \citenamefont
  {Horvath}(1994)}]{creutz1994surface}%
  \BibitemOpen
  \bibfield  {author} {\bibinfo {author} {\bibfnamefont {M.}~\bibnamefont
  {Creutz}}\ and\ \bibinfo {author} {\bibfnamefont {I.}~\bibnamefont
  {Horvath}},\ }\href {\doibase 10.1016/0920-5632(94)90452-9} {\bibfield
  {journal} {\bibinfo  {journal} {Nucl. Phys. B}\ }\textbf {\bibinfo {volume}
  {34}},\ \bibinfo {pages} {583} (\bibinfo {year} {1994})}\BibitemShut
  {NoStop}%
\bibitem [{\citenamefont {Nielsen}\ and\ \citenamefont
  {Ninomiya}(1981{\natexlab{a}})}]{NIELSEN1981219}%
  \BibitemOpen
  \bibfield  {author} {\bibinfo {author} {\bibfnamefont {H.~B.}\ \bibnamefont
  {Nielsen}}\ and\ \bibinfo {author} {\bibfnamefont {M.}~\bibnamefont
  {Ninomiya}},\ }\href {\doibase 10.1016/0370-2693(81)91026-1} {\bibfield
  {journal} {\bibinfo  {journal} {Phys. Lett. B}\ }\textbf {\bibinfo {volume}
  {105}},\ \bibinfo {pages} {219 } (\bibinfo {year}
  {1981}{\natexlab{a}})}\BibitemShut {NoStop}%
\bibitem [{\citenamefont {Nielsen}\ and\ \citenamefont
  {Ninomiya}(1981{\natexlab{b}})}]{nielsen1981absence}%
  \BibitemOpen
  \bibfield  {author} {\bibinfo {author} {\bibfnamefont {H.~B.}\ \bibnamefont
  {Nielsen}}\ and\ \bibinfo {author} {\bibfnamefont {M.}~\bibnamefont
  {Ninomiya}},\ }\href {\doibase 10.1016/0550-3213(81)90361-8} {\bibfield
  {journal} {\bibinfo  {journal} {Nucl. Phys. B}\ }\textbf {\bibinfo {volume}
  {185}},\ \bibinfo {pages} {20} (\bibinfo {year}
  {1981}{\natexlab{b}})}\BibitemShut {NoStop}%
\bibitem [{\citenamefont {Karsten}(1981)}]{Karsten1981}%
  \BibitemOpen
  \bibfield  {author} {\bibinfo {author} {\bibfnamefont {L.~H.}\ \bibnamefont
  {Karsten}},\ }\href {\doibase 10.1016/0370-2693(81)90133-7} {\bibfield
  {journal} {\bibinfo  {journal} {Phys. Lett. B}\ }\textbf {\bibinfo {volume}
  {104}},\ \bibinfo {pages} {315} (\bibinfo {year} {1981})}\BibitemShut
  {NoStop}%
\bibitem [{\citenamefont {Klein}(1929)}]{Klein1929}%
  \BibitemOpen
  \bibfield  {author} {\bibinfo {author} {\bibfnamefont {O.}~\bibnamefont
  {Klein}},\ }\href {\doibase 10.1007/BF01339716} {\bibfield  {journal}
  {\bibinfo  {journal} {Z. Phys.}\ }\textbf {\bibinfo {volume} {53}},\ \bibinfo
  {pages} {157} (\bibinfo {year} {1929})}\BibitemShut {NoStop}%
\bibitem [{\citenamefont {Katsnelson}\ \emph {et~al.}(2006)\citenamefont
  {Katsnelson}, \citenamefont {Novoselov},\ and\ \citenamefont
  {Geim}}]{katsnelson2006chiral}%
  \BibitemOpen
  \bibfield  {author} {\bibinfo {author} {\bibfnamefont {M.}~\bibnamefont
  {Katsnelson}}, \bibinfo {author} {\bibfnamefont {K.}~\bibnamefont
  {Novoselov}}, \ and\ \bibinfo {author} {\bibfnamefont {A.}~\bibnamefont
  {Geim}},\ }\href {\doibase 10.1038/nphys384} {\bibfield  {journal} {\bibinfo
  {journal} {Nat. Phys.}\ }\textbf {\bibinfo {volume} {2}},\ \bibinfo {pages}
  {620} (\bibinfo {year} {2006})}\BibitemShut {NoStop}%
\bibitem [{\citenamefont {Berry}\ and\ \citenamefont
  {Mondragon}(1987)}]{berry1987neutrino}%
  \BibitemOpen
  \bibfield  {author} {\bibinfo {author} {\bibfnamefont {M.~V.}\ \bibnamefont
  {Berry}}\ and\ \bibinfo {author} {\bibfnamefont {R.~J.}\ \bibnamefont
  {Mondragon}},\ }\href {\doibase 10.1098/rspa.1987.0080} {\bibfield  {journal}
  {\bibinfo  {journal} {Proc. R. Soc. Lond., Ser. A}\ }\textbf {\bibinfo
  {volume} {412}},\ \bibinfo {pages} {53} (\bibinfo {year} {1987})}\BibitemShut
  {NoStop}%
\bibitem [{\citenamefont {Alonso}\ \emph {et~al.}(1997)\citenamefont {Alonso},
  \citenamefont {Vincenzo},\ and\ \citenamefont
  {Mondino}}]{Alonso1997Boundary}%
  \BibitemOpen
  \bibfield  {author} {\bibinfo {author} {\bibfnamefont {V.}~\bibnamefont
  {Alonso}}, \bibinfo {author} {\bibfnamefont {S.~D.}\ \bibnamefont
  {Vincenzo}}, \ and\ \bibinfo {author} {\bibfnamefont {L.}~\bibnamefont
  {Mondino}},\ }\href {\doibase 10.1088/0143-0807/18/5/001} {\bibfield
  {journal} {\bibinfo  {journal} {Eur. J. Phys.}\ }\textbf {\bibinfo {volume}
  {18}},\ \bibinfo {pages} {315} (\bibinfo {year} {1997})}\BibitemShut
  {NoStop}%
\bibitem [{\citenamefont {McCann}\ and\ \citenamefont
  {Fal’ko}(2004)}]{mccann2004symmetry}%
  \BibitemOpen
  \bibfield  {author} {\bibinfo {author} {\bibfnamefont {E.}~\bibnamefont
  {McCann}}\ and\ \bibinfo {author} {\bibfnamefont {V.~I.}\ \bibnamefont
  {Fal’ko}},\ }\href {\doibase 10.1088/0953-8984/16/13/016} {\bibfield
  {journal} {\bibinfo  {journal} {J. Phys. Condens. Mat.}\ }\textbf {\bibinfo
  {volume} {16}},\ \bibinfo {pages} {2371} (\bibinfo {year}
  {2004})}\BibitemShut {NoStop}%
\bibitem [{\citenamefont {Ertler}\ \emph {et~al.}(2014)\citenamefont {Ertler},
  \citenamefont {Raith},\ and\ \citenamefont {Fabian}}]{Fabian2014TIDots}%
  \BibitemOpen
  \bibfield  {author} {\bibinfo {author} {\bibfnamefont {C.}~\bibnamefont
  {Ertler}}, \bibinfo {author} {\bibfnamefont {M.}~\bibnamefont {Raith}}, \
  and\ \bibinfo {author} {\bibfnamefont {J.}~\bibnamefont {Fabian}},\ }\href
  {\doibase 10.1103/PhysRevB.89.075432} {\bibfield  {journal} {\bibinfo
  {journal} {Phys. Rev. B}\ }\textbf {\bibinfo {volume} {89}},\ \bibinfo
  {pages} {075432} (\bibinfo {year} {2014})}\BibitemShut {NoStop}%
\bibitem [{\citenamefont {Ferreira}\ and\ \citenamefont
  {Loss}(2013)}]{ferreira2013magnetically}%
  \BibitemOpen
  \bibfield  {author} {\bibinfo {author} {\bibfnamefont {G.~J.}\ \bibnamefont
  {Ferreira}}\ and\ \bibinfo {author} {\bibfnamefont {D.}~\bibnamefont
  {Loss}},\ }\href {\doibase 10.1103/PhysRevLett.111.106802} {\bibfield
  {journal} {\bibinfo  {journal} {Phys. Rev. Lett.}\ }\textbf {\bibinfo
  {volume} {111}},\ \bibinfo {pages} {106802} (\bibinfo {year}
  {2013})}\BibitemShut {NoStop}%
\bibitem [{\citenamefont {Yeats}\ \emph {et~al.}(2017)\citenamefont {Yeats},
  \citenamefont {Mintun}, \citenamefont {Pan}, \citenamefont {Richardella},
  \citenamefont {Buckley}, \citenamefont {Samarth},\ and\ \citenamefont
  {Awschalom}}]{yeats2017local}%
  \BibitemOpen
  \bibfield  {author} {\bibinfo {author} {\bibfnamefont {A.~L.}\ \bibnamefont
  {Yeats}}, \bibinfo {author} {\bibfnamefont {P.~J.}\ \bibnamefont {Mintun}},
  \bibinfo {author} {\bibfnamefont {Y.}~\bibnamefont {Pan}}, \bibinfo {author}
  {\bibfnamefont {A.}~\bibnamefont {Richardella}}, \bibinfo {author}
  {\bibfnamefont {B.~B.}\ \bibnamefont {Buckley}}, \bibinfo {author}
  {\bibfnamefont {N.}~\bibnamefont {Samarth}}, \ and\ \bibinfo {author}
  {\bibfnamefont {D.~D.}\ \bibnamefont {Awschalom}},\ }\href
  {https://doi.org/10.1073/pnas.1713458114} {\bibfield  {journal} {\bibinfo
  {journal} {Proc. Natl. Acad. Sci. U.S.A.}\ }\textbf {\bibinfo {volume} {114}}
  (\bibinfo {year} {2017})}\BibitemShut {NoStop}%
\bibitem [{\citenamefont {Ferreira}\ \emph {et~al.}(2011)\citenamefont
  {Ferreira}, \citenamefont {Leuenberger}, \citenamefont {Loss},\ and\
  \citenamefont {Egues}}]{Ferreira2011LowBiasNDR}%
  \BibitemOpen
  \bibfield  {author} {\bibinfo {author} {\bibfnamefont {G.~J.}\ \bibnamefont
  {Ferreira}}, \bibinfo {author} {\bibfnamefont {M.~N.}\ \bibnamefont
  {Leuenberger}}, \bibinfo {author} {\bibfnamefont {D.}~\bibnamefont {Loss}}, \
  and\ \bibinfo {author} {\bibfnamefont {J.~C.}\ \bibnamefont {Egues}},\ }\href
  {\doibase 10.1103/PhysRevB.84.125453} {\bibfield  {journal} {\bibinfo
  {journal} {Phys. Rev. B}\ }\textbf {\bibinfo {volume} {84}},\ \bibinfo
  {pages} {125453} (\bibinfo {year} {2011})}\BibitemShut {NoStop}%
\bibitem [{\citenamefont {Zhang}\ \emph {et~al.}(2009)\citenamefont {Zhang},
  \citenamefont {Liu}, \citenamefont {Qi}, \citenamefont {Dai}, \citenamefont
  {Fang},\ and\ \citenamefont {Zhang}}]{zhang2009topological}%
  \BibitemOpen
  \bibfield  {author} {\bibinfo {author} {\bibfnamefont {H.}~\bibnamefont
  {Zhang}}, \bibinfo {author} {\bibfnamefont {C.-X.}\ \bibnamefont {Liu}},
  \bibinfo {author} {\bibfnamefont {X.-L.}\ \bibnamefont {Qi}}, \bibinfo
  {author} {\bibfnamefont {X.}~\bibnamefont {Dai}}, \bibinfo {author}
  {\bibfnamefont {Z.}~\bibnamefont {Fang}}, \ and\ \bibinfo {author}
  {\bibfnamefont {S.-C.}\ \bibnamefont {Zhang}},\ }\href {\doibase
  10.1038/NPHYS1270} {\bibfield  {journal} {\bibinfo  {journal} {Nat. Phys.}\
  }\textbf {\bibinfo {volume} {5}},\ \bibinfo {pages} {438} (\bibinfo {year}
  {2009})}\BibitemShut {NoStop}%
\bibitem [{\citenamefont {Shan}\ \emph {et~al.}(2010)\citenamefont {Shan},
  \citenamefont {Lu},\ and\ \citenamefont {Shen}}]{shan2010effective}%
  \BibitemOpen
  \bibfield  {author} {\bibinfo {author} {\bibfnamefont {W.-Y.}\ \bibnamefont
  {Shan}}, \bibinfo {author} {\bibfnamefont {H.-Z.}\ \bibnamefont {Lu}}, \ and\
  \bibinfo {author} {\bibfnamefont {S.-Q.}\ \bibnamefont {Shen}},\ }\href
  {\doibase 10.1088/1367-2630/12/4/043048} {\bibfield  {journal} {\bibinfo
  {journal} {New J. Phys.}\ }\textbf {\bibinfo {volume} {12}},\ \bibinfo
  {pages} {043048} (\bibinfo {year} {2010})}\BibitemShut {NoStop}%
\bibitem [{\citenamefont {Kresse}\ and\ \citenamefont
  {Furthm{\"u}ller}(1996)}]{vasp1}%
  \BibitemOpen
  \bibfield  {author} {\bibinfo {author} {\bibfnamefont {G.}~\bibnamefont
  {Kresse}}\ and\ \bibinfo {author} {\bibfnamefont {J.}~\bibnamefont
  {Furthm{\"u}ller}},\ }\href {\doibase 10.1016/0927-0256(96)00008-0}
  {\bibfield  {journal} {\bibinfo  {journal} {Comput. Mater. Sci.}\ }\textbf
  {\bibinfo {volume} {6}},\ \bibinfo {pages} {15} (\bibinfo {year}
  {1996})}\BibitemShut {NoStop}%
\bibitem [{SM()}]{SM}%
  \BibitemOpen
  \href@noop {} {}\bibinfo {note} {See Supplemental Material at
  \url{http://link.aps.org/supplemental/00.0000/xxxxxxxxxxx.000.000000} for
  additional details, which includes Refs.~\onlinecite{PBE,
  Monkhorst1976BZpoints, Blochl1994PAW, VDW-DF2, PRB80Mei,
  PRB82Zhao}.}\BibitemShut {Stop}%
\bibitem [{\citenamefont {Groth}\ \emph {et~al.}(2014)\citenamefont {Groth},
  \citenamefont {Wimmer}, \citenamefont {Akhmerov},\ and\ \citenamefont
  {Waintal}}]{kwant}%
  \BibitemOpen
  \bibfield  {author} {\bibinfo {author} {\bibfnamefont {C.~W.}\ \bibnamefont
  {Groth}}, \bibinfo {author} {\bibfnamefont {M.}~\bibnamefont {Wimmer}},
  \bibinfo {author} {\bibfnamefont {A.~R.}\ \bibnamefont {Akhmerov}}, \ and\
  \bibinfo {author} {\bibfnamefont {X.}~\bibnamefont {Waintal}},\ }\href
  {\doibase doi:10.1088/1367-2630/16/6/063065} {\bibfield  {journal} {\bibinfo
  {journal} {New J. Phys.}\ }\textbf {\bibinfo {volume} {16}},\ \bibinfo
  {pages} {063065} (\bibinfo {year} {2014})}\BibitemShut {NoStop}%
\bibitem [{\citenamefont {F\"orster}\ \emph {et~al.}(2015)\citenamefont
  {F\"orster}, \citenamefont {Kr\"uger},\ and\ \citenamefont
  {Rohlfing}}]{Forster2015GW}%
  \BibitemOpen
  \bibfield  {author} {\bibinfo {author} {\bibfnamefont {T.}~\bibnamefont
  {F\"orster}}, \bibinfo {author} {\bibfnamefont {P.}~\bibnamefont {Kr\"uger}},
  \ and\ \bibinfo {author} {\bibfnamefont {M.}~\bibnamefont {Rohlfing}},\
  }\href {\doibase 10.1103/PhysRevB.92.201404} {\bibfield  {journal} {\bibinfo
  {journal} {Phys. Rev. B}\ }\textbf {\bibinfo {volume} {92}},\ \bibinfo
  {pages} {201404} (\bibinfo {year} {2015})}\BibitemShut {NoStop}%
\bibitem [{\citenamefont {Zhang}\ \emph {et~al.}(2010)\citenamefont {Zhang},
  \citenamefont {He}, \citenamefont {Chang}, \citenamefont {Song},
  \citenamefont {Wang}, \citenamefont {Chen}, \citenamefont {Jia},
  \citenamefont {Fang}, \citenamefont {Dai}, \citenamefont {Shan},
  \citenamefont {Shen}, \citenamefont {Niu}, \citenamefont {Qi}, \citenamefont
  {Zhang}, \citenamefont {Ma},\ and\ \citenamefont {Xue}}]{natureZhang2010}%
  \BibitemOpen
  \bibfield  {author} {\bibinfo {author} {\bibfnamefont {Y.}~\bibnamefont
  {Zhang}}, \bibinfo {author} {\bibfnamefont {K.}~\bibnamefont {He}}, \bibinfo
  {author} {\bibfnamefont {C.-Z.}\ \bibnamefont {Chang}}, \bibinfo {author}
  {\bibfnamefont {C.-L.}\ \bibnamefont {Song}}, \bibinfo {author}
  {\bibfnamefont {L.-L.}\ \bibnamefont {Wang}}, \bibinfo {author}
  {\bibfnamefont {X.}~\bibnamefont {Chen}}, \bibinfo {author} {\bibfnamefont
  {J.-F.}\ \bibnamefont {Jia}}, \bibinfo {author} {\bibfnamefont
  {Z.}~\bibnamefont {Fang}}, \bibinfo {author} {\bibfnamefont {X.}~\bibnamefont
  {Dai}}, \bibinfo {author} {\bibfnamefont {W.-Y.}\ \bibnamefont {Shan}},
  \bibinfo {author} {\bibfnamefont {S.-Q.}\ \bibnamefont {Shen}}, \bibinfo
  {author} {\bibfnamefont {Q.}~\bibnamefont {Niu}}, \bibinfo {author}
  {\bibfnamefont {X.-L.}\ \bibnamefont {Qi}}, \bibinfo {author} {\bibfnamefont
  {S.-C.}\ \bibnamefont {Zhang}}, \bibinfo {author} {\bibfnamefont {X.-C.}\
  \bibnamefont {Ma}}, \ and\ \bibinfo {author} {\bibfnamefont {Q.-K.}\
  \bibnamefont {Xue}},\ }\href {\doibase 10.1038/nphys1689} {\bibfield
  {journal} {\bibinfo  {journal} {Nat Phys}\ }\textbf {\bibinfo {volume} {6}},\
  \bibinfo {pages} {584} (\bibinfo {year} {2010})}\BibitemShut {NoStop}%
\bibitem [{\citenamefont {Altland}\ and\ \citenamefont
  {Zirnbauer}(1997)}]{PhysRevB.55.1142}%
  \BibitemOpen
  \bibfield  {author} {\bibinfo {author} {\bibfnamefont {A.}~\bibnamefont
  {Altland}}\ and\ \bibinfo {author} {\bibfnamefont {M.~R.}\ \bibnamefont
  {Zirnbauer}},\ }\href {\doibase 10.1103/PhysRevB.55.1142} {\bibfield
  {journal} {\bibinfo  {journal} {Phys. Rev. B}\ }\textbf {\bibinfo {volume}
  {55}},\ \bibinfo {pages} {1142} (\bibinfo {year} {1997})}\BibitemShut
  {NoStop}%
\bibitem [{\citenamefont {Bernard}\ and\ \citenamefont
  {LeClair}(2002)}]{bernard2002classification}%
  \BibitemOpen
  \bibfield  {author} {\bibinfo {author} {\bibfnamefont {D.}~\bibnamefont
  {Bernard}}\ and\ \bibinfo {author} {\bibfnamefont {A.}~\bibnamefont
  {LeClair}},\ }\href {\doibase 10.1088/0305-4470/35/11/303} {\bibfield
  {journal} {\bibinfo  {journal} {J. Phys. A Math. Gen.}\ }\textbf {\bibinfo
  {volume} {35}},\ \bibinfo {pages} {2555} (\bibinfo {year}
  {2002})}\BibitemShut {NoStop}%
\bibitem [{\citenamefont {Schnyder}\ \emph {et~al.}(2008)\citenamefont
  {Schnyder}, \citenamefont {Ryu}, \citenamefont {Furusaki},\ and\
  \citenamefont {Ludwig}}]{Schnyder2008PRBClassification}%
  \BibitemOpen
  \bibfield  {author} {\bibinfo {author} {\bibfnamefont {A.~P.}\ \bibnamefont
  {Schnyder}}, \bibinfo {author} {\bibfnamefont {S.}~\bibnamefont {Ryu}},
  \bibinfo {author} {\bibfnamefont {A.}~\bibnamefont {Furusaki}}, \ and\
  \bibinfo {author} {\bibfnamefont {A.~W.~W.}\ \bibnamefont {Ludwig}},\ }\href
  {\doibase 10.1103/PhysRevB.78.195125} {\bibfield  {journal} {\bibinfo
  {journal} {Phys. Rev. B}\ }\textbf {\bibinfo {volume} {78}},\ \bibinfo
  {pages} {195125} (\bibinfo {year} {2008})}\BibitemShut {NoStop}%
\bibitem [{\citenamefont {Cho}\ \emph {et~al.}(2016)\citenamefont {Cho},
  \citenamefont {Zhong}, \citenamefont {Schneeloch}, \citenamefont {Gu},\ and\
  \citenamefont {Mason}}]{SRSungjae2016}%
  \BibitemOpen
  \bibfield  {author} {\bibinfo {author} {\bibfnamefont {S.}~\bibnamefont
  {Cho}}, \bibinfo {author} {\bibfnamefont {R.}~\bibnamefont {Zhong}}, \bibinfo
  {author} {\bibfnamefont {J.~A.}\ \bibnamefont {Schneeloch}}, \bibinfo
  {author} {\bibfnamefont {G.}~\bibnamefont {Gu}}, \ and\ \bibinfo {author}
  {\bibfnamefont {N.}~\bibnamefont {Mason}},\ }\href {\doibase
  10.1038/srep21767} {\bibfield  {journal} {\bibinfo  {journal} {Sci. Rep.}\
  }\textbf {\bibinfo {volume} {6}},\ \bibinfo {pages} {21767} (\bibinfo {year}
  {2016})}\BibitemShut {NoStop}%
\bibitem [{\citenamefont {de~Oliveira}\ \emph {et~al.}(2017)\citenamefont
  {de~Oliveira}, \citenamefont {Scopel},\ and\ \citenamefont
  {Miwa}}]{JPCMMiwa2017}%
  \BibitemOpen
  \bibfield  {author} {\bibinfo {author} {\bibfnamefont {I.~S.~S.}\
  \bibnamefont {de~Oliveira}}, \bibinfo {author} {\bibfnamefont {W.~L.}\
  \bibnamefont {Scopel}}, \ and\ \bibinfo {author} {\bibfnamefont {R.~H.}\
  \bibnamefont {Miwa}},\ }\href {\doibase 10.1088/1361-648X/29/4/045302}
  {\bibfield  {journal} {\bibinfo  {journal} {J. Phys. Condens. Mat.}\ }\textbf
  {\bibinfo {volume} {29}},\ \bibinfo {pages} {045302} (\bibinfo {year}
  {2017})}\BibitemShut {NoStop}%
\bibitem [{\citenamefont {Nakada}\ \emph {et~al.}(1996)\citenamefont {Nakada},
  \citenamefont {Fujita}, \citenamefont {Dresselhaus},\ and\ \citenamefont
  {Dresselhaus}}]{Nakada1996GrapheneRibbons}%
  \BibitemOpen
  \bibfield  {author} {\bibinfo {author} {\bibfnamefont {K.}~\bibnamefont
  {Nakada}}, \bibinfo {author} {\bibfnamefont {M.}~\bibnamefont {Fujita}},
  \bibinfo {author} {\bibfnamefont {G.}~\bibnamefont {Dresselhaus}}, \ and\
  \bibinfo {author} {\bibfnamefont {M.~S.}\ \bibnamefont {Dresselhaus}},\
  }\href {\doibase 10.1103/PhysRevB.54.17954} {\bibfield  {journal} {\bibinfo
  {journal} {Phys. Rev. B}\ }\textbf {\bibinfo {volume} {54}},\ \bibinfo
  {pages} {17954} (\bibinfo {year} {1996})}\BibitemShut {NoStop}%
\bibitem [{\citenamefont {Brey}\ and\ \citenamefont
  {Fertig}(2006{\natexlab{a}})}]{BreyFertig2006Graphene}%
  \BibitemOpen
  \bibfield  {author} {\bibinfo {author} {\bibfnamefont {L.}~\bibnamefont
  {Brey}}\ and\ \bibinfo {author} {\bibfnamefont {H.~A.}\ \bibnamefont
  {Fertig}},\ }\href {\doibase 10.1103/PhysRevB.73.195408} {\bibfield
  {journal} {\bibinfo  {journal} {Phys. Rev. B}\ }\textbf {\bibinfo {volume}
  {73}},\ \bibinfo {pages} {195408} (\bibinfo {year}
  {2006}{\natexlab{a}})}\BibitemShut {NoStop}%
\bibitem [{\citenamefont {Brey}\ and\ \citenamefont
  {Fertig}(2006{\natexlab{b}})}]{BreyFertig2006Dirac}%
  \BibitemOpen
  \bibfield  {author} {\bibinfo {author} {\bibfnamefont {L.}~\bibnamefont
  {Brey}}\ and\ \bibinfo {author} {\bibfnamefont {H.~A.}\ \bibnamefont
  {Fertig}},\ }\href {\doibase 10.1103/PhysRevB.73.235411} {\bibfield
  {journal} {\bibinfo  {journal} {Phys. Rev. B}\ }\textbf {\bibinfo {volume}
  {73}},\ \bibinfo {pages} {235411} (\bibinfo {year}
  {2006}{\natexlab{b}})}\BibitemShut {NoStop}%
\bibitem [{\citenamefont {Castro~Neto}\ \emph {et~al.}(2009)\citenamefont
  {Castro~Neto}, \citenamefont {Guinea}, \citenamefont {Peres}, \citenamefont
  {Novoselov},\ and\ \citenamefont {Geim}}]{CastroNeto2009Review}%
  \BibitemOpen
  \bibfield  {author} {\bibinfo {author} {\bibfnamefont {A.~H.}\ \bibnamefont
  {Castro~Neto}}, \bibinfo {author} {\bibfnamefont {F.}~\bibnamefont {Guinea}},
  \bibinfo {author} {\bibfnamefont {N.~M.~R.}\ \bibnamefont {Peres}}, \bibinfo
  {author} {\bibfnamefont {K.~S.}\ \bibnamefont {Novoselov}}, \ and\ \bibinfo
  {author} {\bibfnamefont {A.~K.}\ \bibnamefont {Geim}},\ }\href {\doibase
  10.1103/RevModPhys.81.109} {\bibfield  {journal} {\bibinfo  {journal} {Rev.
  Mod. Phys.}\ }\textbf {\bibinfo {volume} {81}},\ \bibinfo {pages} {109}
  (\bibinfo {year} {2009})}\BibitemShut {NoStop}%
\bibitem [{\citenamefont {Bernevig}\ \emph {et~al.}(2006)\citenamefont
  {Bernevig}, \citenamefont {Hughes},\ and\ \citenamefont
  {Zhang}}]{bernevig2006quantum}%
  \BibitemOpen
  \bibfield  {author} {\bibinfo {author} {\bibfnamefont {B.~A.}\ \bibnamefont
  {Bernevig}}, \bibinfo {author} {\bibfnamefont {T.~L.}\ \bibnamefont
  {Hughes}}, \ and\ \bibinfo {author} {\bibfnamefont {S.-C.}\ \bibnamefont
  {Zhang}},\ }\href {\doibase 10.1126/science.1133734} {\bibfield  {journal}
  {\bibinfo  {journal} {Science}\ }\textbf {\bibinfo {volume} {314}},\ \bibinfo
  {pages} {1757} (\bibinfo {year} {2006})}\BibitemShut {NoStop}%
\bibitem [{\citenamefont {Michetti}\ \emph {et~al.}(2012)\citenamefont
  {Michetti}, \citenamefont {Penteado}, \citenamefont {Egues},\ and\
  \citenamefont {Recher}}]{michetti2012helical}%
  \BibitemOpen
  \bibfield  {author} {\bibinfo {author} {\bibfnamefont {P.}~\bibnamefont
  {Michetti}}, \bibinfo {author} {\bibfnamefont {P.}~\bibnamefont {Penteado}},
  \bibinfo {author} {\bibfnamefont {J.~C.}\ \bibnamefont {Egues}}, \ and\
  \bibinfo {author} {\bibfnamefont {P.}~\bibnamefont {Recher}},\ }\href
  {\doibase 10.1088/0268-1242/27/12/124007} {\bibfield  {journal} {\bibinfo
  {journal} {Semicond. Sci. Technol.}\ }\textbf {\bibinfo {volume} {27}},\
  \bibinfo {pages} {124007} (\bibinfo {year} {2012})}\BibitemShut {NoStop}%
\bibitem [{\citenamefont {Perdew}\ \emph {et~al.}(1996)\citenamefont {Perdew},
  \citenamefont {Burke},\ and\ \citenamefont {Ernzerhof}}]{PBE}%
  \BibitemOpen
  \bibfield  {author} {\bibinfo {author} {\bibfnamefont {J.~P.}\ \bibnamefont
  {Perdew}}, \bibinfo {author} {\bibfnamefont {K.}~\bibnamefont {Burke}}, \
  and\ \bibinfo {author} {\bibfnamefont {M.}~\bibnamefont {Ernzerhof}},\ }\href
  {\doibase 10.1103/PhysRevLett.77.3865} {\bibfield  {journal} {\bibinfo
  {journal} {Phys. Rev. Lett.}\ }\textbf {\bibinfo {volume} {77}},\ \bibinfo
  {pages} {3865} (\bibinfo {year} {1996})}\BibitemShut {NoStop}%
\bibitem [{\citenamefont {Monkhorst}\ and\ \citenamefont
  {Pack}(1976)}]{Monkhorst1976BZpoints}%
  \BibitemOpen
  \bibfield  {author} {\bibinfo {author} {\bibfnamefont {H.~J.}\ \bibnamefont
  {Monkhorst}}\ and\ \bibinfo {author} {\bibfnamefont {J.~D.}\ \bibnamefont
  {Pack}},\ }\href {\doibase 10.1103/PhysRevB.13.5188} {\bibfield  {journal}
  {\bibinfo  {journal} {Phys. Rev. B}\ }\textbf {\bibinfo {volume} {13}},\
  \bibinfo {pages} {5188} (\bibinfo {year} {1976})}\BibitemShut {NoStop}%
\bibitem [{\citenamefont {Bl\"ochl}(1994)}]{Blochl1994PAW}%
  \BibitemOpen
  \bibfield  {author} {\bibinfo {author} {\bibfnamefont {P.~E.}\ \bibnamefont
  {Bl\"ochl}},\ }\href {\doibase 10.1103/PhysRevB.50.17953} {\bibfield
  {journal} {\bibinfo  {journal} {Phys. Rev. B}\ }\textbf {\bibinfo {volume}
  {50}},\ \bibinfo {pages} {17953} (\bibinfo {year} {1994})}\BibitemShut
  {NoStop}%
\bibitem [{\citenamefont {Lee}\ \emph {et~al.}(2010)\citenamefont {Lee},
  \citenamefont {Murray}, \citenamefont {Kong}, \citenamefont {Lundqvist},\
  and\ \citenamefont {Langreth}}]{VDW-DF2}%
  \BibitemOpen
  \bibfield  {author} {\bibinfo {author} {\bibfnamefont {K.}~\bibnamefont
  {Lee}}, \bibinfo {author} {\bibfnamefont {E.~D.}\ \bibnamefont {Murray}},
  \bibinfo {author} {\bibfnamefont {L.}~\bibnamefont {Kong}}, \bibinfo {author}
  {\bibfnamefont {B.~I.}\ \bibnamefont {Lundqvist}}, \ and\ \bibinfo {author}
  {\bibfnamefont {D.~C.}\ \bibnamefont {Langreth}},\ }\href {\doibase
  10.1103/PhysRevB.82.081101} {\bibfield  {journal} {\bibinfo  {journal} {Phys.
  Rev. B}\ }\textbf {\bibinfo {volume} {82}},\ \bibinfo {pages} {081101}
  (\bibinfo {year} {2010})}\BibitemShut {NoStop}%
\bibitem [{\citenamefont {Mei}\ \emph {et~al.}(2009)\citenamefont {Mei},
  \citenamefont {Shang}, \citenamefont {Wang},\ and\ \citenamefont
  {Liu}}]{PRB80Mei}%
  \BibitemOpen
  \bibfield  {author} {\bibinfo {author} {\bibfnamefont {Z.-G.}\ \bibnamefont
  {Mei}}, \bibinfo {author} {\bibfnamefont {S.-L.}\ \bibnamefont {Shang}},
  \bibinfo {author} {\bibfnamefont {Y.}~\bibnamefont {Wang}}, \ and\ \bibinfo
  {author} {\bibfnamefont {Z.-K.}\ \bibnamefont {Liu}},\ }\href {\doibase
  10.1103/PhysRevB.80.104116} {\bibfield  {journal} {\bibinfo  {journal} {Phys.
  Rev. B}\ }\textbf {\bibinfo {volume} {80}},\ \bibinfo {pages} {104116}
  (\bibinfo {year} {2009})}\BibitemShut {NoStop}%
\bibitem [{\citenamefont {Zhao}\ \emph {et~al.}(2010)\citenamefont {Zhao},
  \citenamefont {Zhang},\ and\ \citenamefont {Lababidi}}]{PRB82Zhao}%
  \BibitemOpen
  \bibfield  {author} {\bibinfo {author} {\bibfnamefont {E.}~\bibnamefont
  {Zhao}}, \bibinfo {author} {\bibfnamefont {C.}~\bibnamefont {Zhang}}, \ and\
  \bibinfo {author} {\bibfnamefont {M.}~\bibnamefont {Lababidi}},\ }\href
  {\doibase 10.1103/PhysRevB.82.205331} {\bibfield  {journal} {\bibinfo
  {journal} {Phys. Rev. B}\ }\textbf {\bibinfo {volume} {82}},\ \bibinfo
  {pages} {205331} (\bibinfo {year} {2010})}\BibitemShut {NoStop}%
\end{thebibliography}%

\end{document}